\documentclass[trackchanges,twocolumn,times]{aastex631}

\shorttitle{Plasma composition from SPICE \& EIS}                 

\shortauthors{Brooks et al.}
\journalinfo{To be submitted to the Astrophysical Journal}

\usepackage{grffile}

\begin{document}

\title{Plasma composition measurements in an active region from Solar Orbiter/SPICE and Hinode/EIS}

\author[0000-0002-2189-9313]{David H.\ Brooks}
\affil{College of Science, George Mason University, 4400 University Drive, Fairfax, VA 22030, USA}

\author[0000-0002-6203-5239]{Miho Janvier}      
\affil{Universit\'{e} Paris-Saclay, CNRS, Institut d’Astrophysique Spatiale, 91405, Orsay, France}

\author[0000-0002-0665-2355]{Deborah Baker}    
\affil{Mullard Space Science Laboratory, University College London, Holmbury St. Mary, Dorking, Surrey, RH5 6NT, UK}

\author[0000-0001-6102-6851]{Harry P. Warren}
\affil{Space Science Division, Naval Research Laboratory, Washington, DC 20375, USA}

\author[0000-0003-0972-7022]{Fr\'{e}d\'{e}ric Auch\`{e}re}
\affil{Universit\'{e} Paris-Saclay, CNRS, Institut d’Astrophysique Spatiale, 91405, Orsay, France}

\author[0000-0001-9218-3139]{Mats Carlsson} 
\affil{Institute of Theoretical Astrophysics, University of Oslo, Norway}

\author[0000-0002-6093-7861]{Andrzej Fludra} 
\affil{RAL Space, UKRI STFC Rutherford Appleton Laboratory, Didcot, United Kingdom}

\author[0000-0000-0000-0000]{Don Hassler} 
\affil{Southwest Research Institute, Boulder, CO, USA}

\author[0000-0001-9921-0937]{Hardi Peter} 
\affil{Max-Planck-Institut f\"{u}r Sonnensystemforschung, G\"{o}ttingen, Germany}

\author[0000-0001-9027-9954]{Daniel Müller} 
\affil{European Space Agency, ESTEC, Noordwijk, The Netherlands}

\author[0000-0000-0000-0000]{David Williams} 
\affil{European Space Agency, ESAC, Villanueva de la Ca\~{n}ada, Spain}

\author[0000-0003-1294-1257]{Regina Aznar Cuadrado} 
\affil{Max-Planck-Institut f\"{u}r Sonnensystemforschung, G\"{o}ttingen, Germany}

\author[0000-0001-7090-6180]{Krzysztof Barczynski} 
\affil{Physikalisch-Meteorologisches Observatorium Davos, World Radiation Center, Davos Dorf, Switzerland}
\affil{ETH Z\"{u}rich, IPA, HIT building, Wolfgang-Pauli-Str. 27, 8093 Z\"{u}rich, Switzerland}

\author[0000-0003-4290-1897]{Eric Buchlin} 
\affil{Universit\'{e} Paris-Saclay, CNRS, Institut d’Astrophysique Spatiale, 91405, Orsay, France}

\author[0000-0000-0000-0000]{Martin Caldwell} 
\affil{RAL Space, UKRI STFC Rutherford Appleton Laboratory, Didcot, United Kingdom}

\author[0000-0002-8673-3920]{Terje Fredvik} 
\affil{Institute of Theoretical Astrophysics, University of Oslo, Norway}

\author[0000-0000-0000-0000]{Alessandra Giunta} 
\affil{RAL Space, UKRI STFC Rutherford Appleton Laboratory, Didcot, United Kingdom}

\author[0000-0000-0000-0000]{Tim Grundy} 
\affil{RAL Space, UKRI STFC Rutherford Appleton Laboratory, Didcot, United Kingdom}

\author[0000-0000-0000-0000]{Steve Guest} 
\affil{RAL Space, UKRI STFC Rutherford Appleton Laboratory, Didcot, United Kingdom}

\author[0000-0001-8007-9764]{Margit Haberreiter} 
\affil{Physikalisch-Meteorologisches Observatorium Davos, World Radiation Center, Davos Dorf, Switzerland}

\author[0000-0001-9457-6200]{Louise Harra} 
\affil{Physikalisch-Meteorologisches Observatorium Davos, World Radiation Center, Davos Dorf, Switzerland}
\affil{ETH Z\"{u}rich, IPA, HIT building, Wolfgang-Pauli-Str. 27, 8093 Z\"{u}rich, Switzerland}

\author[0000-0000-0000-0000]{Sarah Leeks} 
\affil{RAL Space, UKRI STFC Rutherford Appleton Laboratory, Didcot, United Kingdom}

\author[0000-0003-1438-1310]{Susanna Parenti} 
\affil{Universit\'{e} Paris-Saclay, CNRS, Institut d’Astrophysique Spatiale, 91405, Orsay, France}

\author[0000-0002-0397-2214]{Gabriel Pelouze} 
\affil{Universit\'{e} Paris-Saclay, CNRS, Institut d’Astrophysique Spatiale, 91405, Orsay, France}

\author[0000-0000-0000-0000]{Joseph Plowman} 
\affil{Southwest Research Institute, Boulder, CO, USA}

\author[0000-0003-1159-5639]{Werner Schmutz} 
\affil{Physikalisch-Meteorologisches Observatorium Davos, World Radiation Center, Davos Dorf, Switzerland}

\author[0000-0000-0000-0000]{Udo Schuehle} 
\affil{Max-Planck-Institut f\"{u}r Sonnensystemforschung, G\"{o}ttingen, Germany}

\author[0000-0000-0000-0000]{Sunil Sidher} 
\affil{RAL Space, UKRI STFC Rutherford Appleton Laboratory, Didcot, United Kingdom}

\author[0000-0001-7298-2320]{Luca Teriaca} 
\affil{Max-Planck-Institut f\"{u}r Sonnensystemforschung, G\"{o}ttingen, Germany}

\author[0000-0002-6895-6426]{William T. Thompson} 
\affil{ADNET Systems Inc., NASA Goddard Space Flight Center, Greenbelt, MD, USA}

\author[0000-0001-9034-2925]{Peter R. Young}
\affil{NASA Goddard Space Flight Center, Greenbelt, MD, USA}
\affil{Department of Mathematics, Physics and Electrical Engineering, Northumbria University, Newcastle upon Tyne, UK}

\begin{abstract}
A key goal of the Solar Orbiter mission is to connect elemental abundance measurements of the solar wind enveloping the spacecraft with EUV spectroscopic observations of their solar sources, but this is not an easy exercise. Observations from previous missions have revealed a highly complex picture of spatial and temporal variations of elemental abundances in the solar corona. We have used coordinated observations from Hinode and Solar Orbiter to attempt new abundance measurements with the SPICE (Spectral Imaging of the Coronal Environment) instrument, and benchmark them against standard analyses from EIS (EUV Imaging Spectrometer). We use observations of several solar features in AR 12781 taken from an Earth-facing view by EIS on 2020 November 10, and SPICE data obtained one week later on 2020 November 17; when the AR had rotated into the Solar Orbiter field-of-view. We identify a range of spectral lines that are useful for determining the transition region and low coronal temperature structure with SPICE, and demonstrate that SPICE measurements are able to differentiate between photospheric and coronal Mg/Ne abundances. The combination of SPICE and EIS is able to establish the atmospheric composition structure of a fan loop/outflow area at the active region edge. We also discuss the problem of resolving the degree of elemental fractionation with SPICE, which is more challenging without further constraints on the temperature structure, and comment on what that can tell us about the sources of the solar wind and solar energetic particles.
\end{abstract}

\section{Introduction}

One of the most important goals of solar-terrestrial physics is to understand how the Sun generates and controls the heliosphere. There are several
open questions that need to be focused on to fully address this topic. For example, what are the sources of the solar wind? Where and how are
solar energetic particles (SEPs) accelerated? How does plasma propagate through the heliosphere? And how does all this information and understanding
feed into a future space weather prediction capability?

With the launch of the Parker Solar Probe \citep[PSP,][]{Fox2016} in 2018 August, and Solar Orbiter in 2020 February \citep{Muller2020}, we now have new capabilities and instrumentation that can be brought to
bear on these questions. PSP and Solar Orbiter will travel to within 0.046 and 0.26\,AU, respectively, at their closest perihelia, while Solar Orbiter will also
rise out of the plane of the ecliptic to observe the polar regions from up to $\sim$33$^\circ$ heliolatitude (in the extended mission phase). Their new observations are already challenging our previous understanding of heliospheric phenomena.
A new question, perhaps, is whether properties and signatures of the solar wind, or SEPs, imprinted in the solar atmosphere and measured and analyzed close to the Sun, survive
propagation through the heliosphere all the way to the Earth. PSP has revealed a complex and dynamic near-Sun magnetic environment replete
with field reversals, for example, see \cite{Bale2019}. The ability to make measurements close-in and far-out from the Sun during the orbits of these missions will help to
complete our understanding of the solar wind, SEPs, and how eruptive phenomena propagate through the heliosphere \citep{Janvier2019}. 

In recent years, a great deal of work has been carried out with the Hinode spacecraft \citep{Kosugi2007}, 
particularly the EUV Imaging Spectrometer \citep[EIS,][]{Culhane2007}, and several diagnostics have emerged. Techniques have been developed to utilize specific element pairs, e.g. Si/S, to trace the source regions \citep{Brooks2011}, 
and indeed, properties of spectra such as line profile asymmetries have been found to be helpful since lines of different elements sometimes show asymmetries of different magnitudes \citep{Brooks2021a}. Of course
elemental abundances are a key tracer for connection science, but studies of solar atmospheric composition are highly complex. The solar corona shows strong spatial
and temporal variability in plasma composition \citep{Feldman1992}. Elements with a low FIP (first ionization potential; $<$ 10\,eV)  are enhanced in the corona by factors of 2--4 \citep[FIP,][]{Pottasch1963,Meyer1985}, but differences are observed between and within long lived structures at the active region boundary and within the core, and changes from coronal composition are seen during impulsive events and flares \citep{Baker2013,Warren2014a,Doschek2015,Warren2016}; see also the review by \cite{DelZanna2018}. 
The situation in the solar wind adds further complexity, with differences in composition seen between fast and slow wind streams, substantial changes within the
slow wind itself, and variability between and within different element ratios \citep{vonSteiger2000,Zurbuchen2002,Zurbuchen2012}.

Nevertheless, the FIP effect has substantial
potential for allowing us to link remote sensing spectroscopic observations with in-situ particle measurements.
Despite variations, the general compositional character of active regions (ARs) is fairly well known, with closed field loops showing a FIP effect that is strongest in the core. This
allows us to link atmospheric and heliospheric phenomena. For example, interplanetary coronal mass ejections (ICMEs) with elevated charge states (indicating higher temperatures) show a stronger FIP effect than is seen in the slow solar
wind \citep{Zurbuchen2016}, suggesting that while the slow wind might emerge from closed field loops, these ICMEs may have been ejected from the AR core. Additionally, photospheric composition is sometimes measured in the high density cores of ICMEs with persistently low charge states \citep{Lepri2021,Rivera2022}, implying that prominence material is being detected.
Hinode observations have also identified previously unnoticed potential sources of the solar wind, such as outflows
at the edges of active regions \citep{Sakao2007,DelZanna2008,Doschek2008,Harra2008}, and these have been linked to the slow solar wind using cross-mission composition measurements between
Hinode and the ACE observatory \citep{Brooks2011}. Although such multi-spacecraft studies have limitations, they have developed the preferred pathway for connection studies with Solar Orbiter, and
built a foundation for the spectroscopic analysis from the Spectral Imaging of the Coronal Environment \citep[SPICE,][]{SPICE2020} instrument.

SPICE itself observes a wavelength range that contains a wide range of emission lines from multiple elements,
and has been designed in part to link with the Solar Wind Analyzer \citep[SWA,][]{Owen2020} for connection science. 
The combination of the Solar Orbiter instruments, together with coordinated observations from EIS and other missions, promises to advance our understanding
of this key topic. Simultaneous observations with SPICE and EIS will rely on a detailed understanding of the cross-instrument calibration.
Of course the first step in making progress is to understand the SPICE
instrument itself, and test and benchmark the 
diagnostics capabilities. This is the subject of this paper. We use observations of AR 12781 on the Earth facing disk from Hinode, together with SPICE observations 
of the same region one week later from the short-term planning (STP-122) commissioning phase. Although the EIS and SPICE observations were not taken
simultaneously,
this allows us to apply well tested and standard elemental abundance measurement routines developed for EIS to understand the compositional
structure of AR 12781, and apply that knowledge to the interpretation of the SPICE observations. 
Based on previous work with similar EIS observations of active regions we expect the strongest FIP-bias to
be found at the loop footpoints \citep{Baker2013,Brooks2021b}. These results suggest a link
with solar energetic particles \citep{Brooks2021b}, which show a higher FIP bias. Generally speaking, upflows in EIS have tended to show a coronal
composition when measured using a Si/S ratio, though a photospheric composition is sometimes seen \citep{Brooks2020}. This is consistent with the slow solar wind, which shows variability
from photospheric to predominantly coronal composition in Si/S measurements. Of course AR structures will evolve over a one week period, but our expectation
is that the general characteristics of AR 12781 should be broadly maintained i.e. whether the bright fan loops show photospheric or coronal composition, and that
similar results should be found with SPICE, using different elements, for this same active region. Furthermore, not all Solar Orbiter encounters will be on 
the Sun-Earth line, so there is value in exploring what can be achieved through multi-viewpoint observations, and through AR tracking on timescales longer than that taken
to cross the disk.
In the process of verification, we identify new methodology for determining elemental abundances and interpreting the new measurements from SPICE. 

\section{Data reduction}
Details of the EIS and SPICE instruments are available in \cite{Culhane2007} and \cite{SPICE2020}. Here we describe the observations 
we use and some of the pertinent characteristics and data reduction methods.

The Hinode/EIS data we use were taken on 2020, November 10, at 02:08:11\,UT\, when the target active region (AR 12781) was
close to disk center from an Earth view. EIS was constructed with an exchangeable slit-assembly 
that can switch between 1$''$, 2$''$, 40$''$, and 266$''$ apertures. These data were taken using the 2$''$ slit. Spectra can be
obtained in two wavelength ranges from 171--211\,\AA\, and 245--291\,\AA\, with a spectral resolution of 23\,m\AA. 
Generally, a subset of these ranges is telemetered to the
ground, and the observing program we analyze comprised 25 wavelength windows covering a wide selection of spectral lines from
ions of \ion{Mg}{5}--\ion{Mg}{7}, \ion{Fe}{8}--\ion{Fe}{17}, \ion{Si}{10}, \ion{S}{10}, \ion{Ca}{14}--\ion{Ca}{17}, and some higher
temperature \ion{Fe}{24} lines seen during flares. The observing program scans an area of 261$'' \times$512$''$ with coarse
3$''$ steps to reduce the duration to about 1 hour with an exposure time of 40\,s.

The data were processed using standard procedures that are available in SolarSoft \citep[SSW,][]{Freeland1998}. The routine eis\_prep
converts the recorded count rates to physical units (erg cm$^{-2}$ s$^{-1}$ sr$^{-1}$ \AA$^{-1}$) after cleaning the data arrays
of cosmic ray strikes, dusty, warm, and hot pixels, and removal of the dark current pedestal. We used the updated on-orbit
photometric calibration of \cite{Warren2014b} in this work. We assume an uncertainty of 23\% for the line intensities \citep{Lang2006}.
A known issue is that the EIS
spectra drift back and forth across the CCDs (charge-coupled device) due to temperature changes and spacecraft motion around 
Hinode's polar orbit. We make an initial correction for this effect using the standard neural network (ANN) model developed by 
\cite{Kamio2010}. This method also corrects the spectral curvature and spatial offsets between the two CCDs. As discussed
elsewhere \citep{Brooks2020}, this ANN model was developed using data early in the mission. As
a result, a residual orbital variation in the spectra is often seen in recent data and is present in our dataset. We removed
this trend using the method discussed by \cite{Brooks2020}. This method involves modeling the residual variation by averaging 
data in the Y-direction over some portion of the field-of-view (FOV). We used the lower 50 pixels for this correction. 
Note that to determine the velocity calibration we use the reference wavelengths derived for the EIS spectral lines by \cite{Warren2011} 
using off-limb quiet Sun observations. All measurements in this paper refer to relative, not absolute, Doppler velocities.

The Solar Orbiter/SPICE data we use were taken 7 days after the EIS observations on 2020, November 17, at 22:28:26\,UT. 
At this time, Solar Orbiter was positioned at 0.91--0.93\,a.u., around 120$^{\circ}$ away from Earth (counter-clockwise behind the Sun), so that
AR 12781 had entered the SPICE FOV. Additional telemetry allowed for science observations to be made for several days around this time-period 
as part of the STP-122 commissioning phase. The SPICE
instrument also operates with four slit width options: 2$''$, 4$''$, 6$''$, and 30$''$. The observations discussed here 
were taken using the 6$''$ slit. The spatial resolution of these SPICE observations is therefore about a factor of 2 lower 
than the EIS observations discussed
here (using the 2$''$ slit and coarse 3$''$ steps). Spectra can be
obtained in two wavelength ranges from 704--790\,\AA\, and 973--1049\,\AA.
The observing program we analyze recorded full detector spectra in these wavelength ranges allowing us to examine a wide selection of spectral lines from
ions of \ion{O}{2}--\ion{O}{6}, \ion{N}{3}--\ion{N}{4}, \ion{Ne}{6}, \ion{Ne}{8}, \ion{C}{3}, \ion{S}{4}--\ion{S}{5}, and \ion{Mg}{8}--\ion{Mg}{9}.
The SPICE observation ID for these data is 33554573 and the observing sequence is a series of single exposures scanning an area of 180$'' \times$1024$''$ in around 50\,min\, with an exposure time of 59.6\,s at each position. The unique identifiers for the exposures we used are V06\_33554573\_000--V06\_33554573\_029 and the data were from a pre-release test. Updated versions of the same files have been released as V08/V09 by the SPICE team  \citep{https://doi.org/10.48326/idoc.medoc.spice.1.0}. 
We used the SUMER spectral atlas of solar-disk features \citep{Curdt2001} to identify the SPICE spectral lines and assign reference wavelengths
that we used to derive velocities.

Several characteristics of the on-orbit performance of SPICE are still being investigated. Preliminary discussions can be
found in the early results article by \cite{Fludra2021}. The commissioning data were calibrated to level-2 and provided
to us for individual testing prior to public release. The data reduction followed developing standard procedures that will
be provided in SSW by the SPICE team. These procedures include all the calibration parameters
that have been quantified to date. The units of the data are W m$^{-2}$ sr$^{-1}$ nm$^{-1}$. We assume an uncertainty of 25\% for
the SPICE line intensities, and this is added in quadrature to the errors from the line fits. For both SPICE, and EIS, the 
line intensities were obtained by fitting Gaussian functions to the spectra. Single or multiple Gaussians were used as appropriate for the
specific line and any associated blends.

\begin{figure}[h]
\centering
\includegraphics[width=0.5\textwidth]{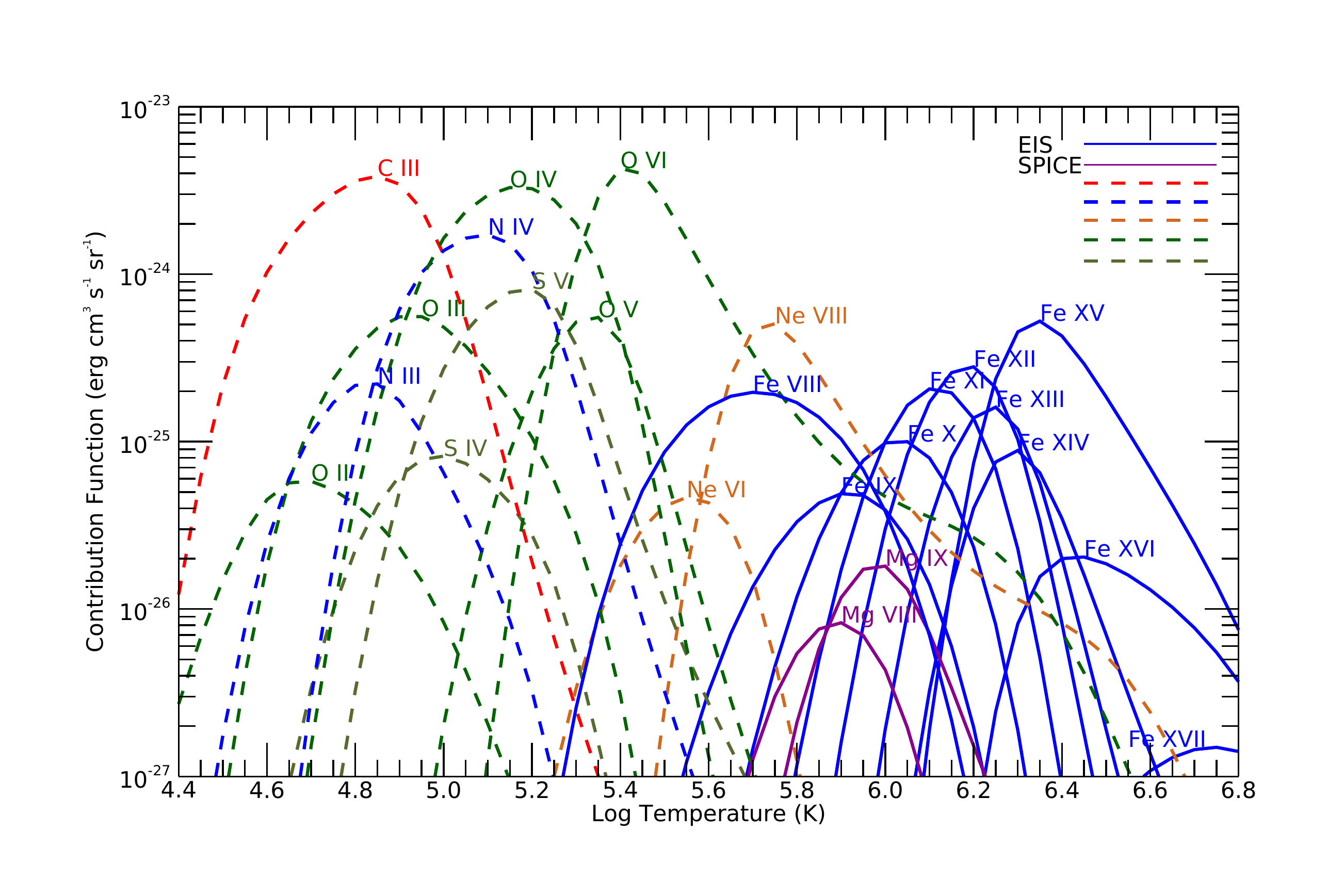}
\caption{ Contribution functions ($G(T,n)$) for many of the SPICE and EIS emission lines analyzed in this work. The plot shows the range of temperature covered by the lines. The calculations assume a constant density (n) of 10$^9$\,cm$^{-3}$ and the adoption of the photospheric abundances of \cite{Scott2015a} and \cite{Scott2015b}. Contribution functions for the elements with FIP $>$10\,eV\, are shown with dashed lines. We only show the strongest line from each ionization stage for clarity. The legend indicates whether the lines are observed by EIS or SPICE. The figure shows that many of the lower temperature high-FIP lines should be relatively bright compared to the higher temperature low-FIP lines
in photospheric conditions. In some regions of the corona, the low-FIP Mg and Fe lines become stronger relative to their photospheric magnitudes.
}
\label{fig0}
\end{figure}

\section{Atomic data and analysis methods} 
\label{adam}
For an optically thin atomic transition from level $i$ to level $j$ the emission line intensity is 
\begin{equation}
I_{i\rightarrow j} = A \int_T G(T,n) \phi (T) dT
\label{eq1}
\end{equation}
where $A$ is the elemental abundance, $T$ is the electron temperature, $n$ is the electron density, $G(T,n)$ is
the contribution function, and $\phi (T)$ is the differential emission measure (DEM). The DEM is a measure of the 
amount of material as a function of temperature. The form of Equation \ref{eq1} assumes a relationship between 
temperature and density, such as constant pressure \citep{Craig1976}. 

The contribution function, $G(T,n)$, theoretically describes the line emission through a population structure 
and ionization equilibrium calculation. These are computed from the relevant electron collisional excitation/deexcitation rates,
spontaneous radiative decay probabilities, and ionization and recombination coefficients. 
For all the spectral lines studied in this work we take the electron collisional and radiative decay data from the CHIANTI 
database \citep{Dere1997} version 10 \citep{DelZanna2021}.

The $G(T,n)$ function is strongly peaked in temperature and weakly sensitive to density; primarily through the suppression of 
dielectronic recombination (DR) at high densities \citep{Burgess1964}, but also due to the effects of metastable levels, and
step-wise (level stepping) ionization and
auto-ionization \citep{Summers2006}. There have been occasional efforts to examine these effects in studies of the solar
corona \citep{Brooks1998,Lanzafame2002,Brooks2006}, but often the densities are too low to show dramatic results. Conversely,
the IRIS instrument observes the lower transition region and chromosphere where densities are higher, so there has been a
renewed interest in including this effect \citep{Young2018}. SPICE observations span
an intermediate temperature and density regime where improved spectroscopic accuracy may also be important \citep{Parenti2019}.
Here we adopt the density dependent ionization equilibrium calculations from the generalized collisional-radiative models 
of ADAS \citep{Summers2006} for 
the elements C, N, O, Ne, and Si, and supplement them with calculations for Mg, S, and Fe using the CHIANTI approximate DR suppression models of 
\cite{Nikolic2018}. 

Our plasma composition measurement technique for EIS is now fairly well established \citep{Brooks2011}. We select a series
of spectral lines from \ion{Fe}{8}--\ion{Fe}{17} for the DEM analysis (Equation \ref{eq1}). We then use the 
\ion{Fe}{13} 202.044/203.826 diagnostic ratio to compute the electron density, and calculate the $G(T,n)$ function for
all the spectral lines adopting the photospheric elemental abundances of \cite{Scott2015a} and \cite{Scott2015b}. 
The DEM is computed at this constant density using the Markov-Chain Monte Carlo (MCMC) algorithm
included in the PintOfAle software package \citep{Kashyap1998,Kashyap2000}. The MCMC algorithm constructs the DEM by 
performing 100 simulations that minimize 
the differences between the observed and computed line intensities and converge to a best-fit solution. Once the DEM is established from the low-FIP Fe
lines, we use it as the basis for computing the FIP bias from the \ion{Si}{10} 258.375\,\AA\, to \ion{S}{10} 264.223\,\AA\,
ratio. Since Si is a low-FIP element, the 258.375\,\AA\, intensity should be well represented by the Fe-only DEM. We check,
however, that there is no large systematic difference, and adjust the DEM to match the \ion{Si}{10} 258.375\,\AA\, intensity.
We then use the adjusted DEM to predict the \ion{S}{10} 264.223\,\AA\, intensity. 
The ratio of the DEM-calculated intensity to the
observed intensity gives us the FIP bias because the \ion{S}{10} 264.223\,\AA\, intensity predicted by the
DEM from the low-FIP elements will be too large/correct depending on whether the actual abundances are coronal/photospheric.
S lies on the boundary
between low- and high-FIP elements and shows variable behavior that is sometimes consistent with one group or the other.
In some theoretical models this depends on the open or closed topology of the magnetic field \citep{Laming2019,Kuroda2020}, making it 
a useful diagnostic regardless \citep{Brooks2021b}. 

For SPICE, our purpose is to develop a new useful technique and benchmark it against the EIS results. This task is challenging for
two reasons. First, the SPICE temperature coverage does not extend far beyond 1\,MK\, in non-flaring coronal conditions. This can
be seen from a plot of the $G(T,n)$ functions for the lines investigated in this work in Figure \ref{fig0}. 
The figure also shows the $G(T,n)$ functions for the EIS lines for comparison. 
The highest temperature
SPICE lines from \ion{Mg}{9} reach about 1\,MK, and these are also the
boundary constraints for the DEM analysis which makes matters more difficult;
though higher temperature lines may be detectable in second order in other datasets with longer
exposures, or in flares \citep{SPICE2020}.
Some studies
of full-disk observations of the Sun suggest that the FIP effect is only detectable above 1\,MK\, \citep{Laming1995}, which
puts the highest temperature SPICE lines at the limit of where the effect would be detectable. This could be problematic
for studies of spectra averaged over large spatial scales and deserves our attention.
Of course most SPICE observations will concentrate on specific features. Several previous analyses
have shown that spikey structures at the edges of active regions - more recently named fan loops - can show
a strong FIP effect in the upper transition region when measured using Mg/Ne ratios \citep{Sheeley1996,Young1997,Widing2001} or Fe/O \citep{Warren2016}. 
These results are relevant to our analysis in section \ref{discussion}. 
Furthermore, there are some indications from full-disk 
SDO/EVE spectra that the FIP effect can be detected at somewhat lower temperatures \citep[0.5\,MK,][]{Brooks2017}.
Secondly, the best spectral lines from low- and high-FIP
elements at the high temperature end of the SPICE capabilities come from \ion{Mg}{8}--\ion{Mg}{9} and \ion{Ne}{8}. 
These ions do not have a good
overlap in temperature, and \ion{Ne}{8} is a Li-like ion so has a large high temperature tail that can influence      
the measured FIP bias if it is not modeled correctly. 
There are two \ion{Ne}{8} lines and five \ion{Mg}{8}--\ion{Mg}{9} lines, however, which
should provide good constraints, and additional lines from lower temperature ions of \ion{S}{4} and \ion{S}{5}. It may also be possible to develop a measurement technique from linear combinations of 
spectral lines; such as has been developed for EIS by \cite{Zambrana2019}. \cite{Parenti2021} have also discussed using composition data to link spacecraft measurements
in the solar wind to their sources, and specifically compare results from the \cite{Zambrana2019} method to a DEM technique that is comparable to what we have used here.

\begin{figure}[h]
\centering
\includegraphics[width=0.5\textwidth]{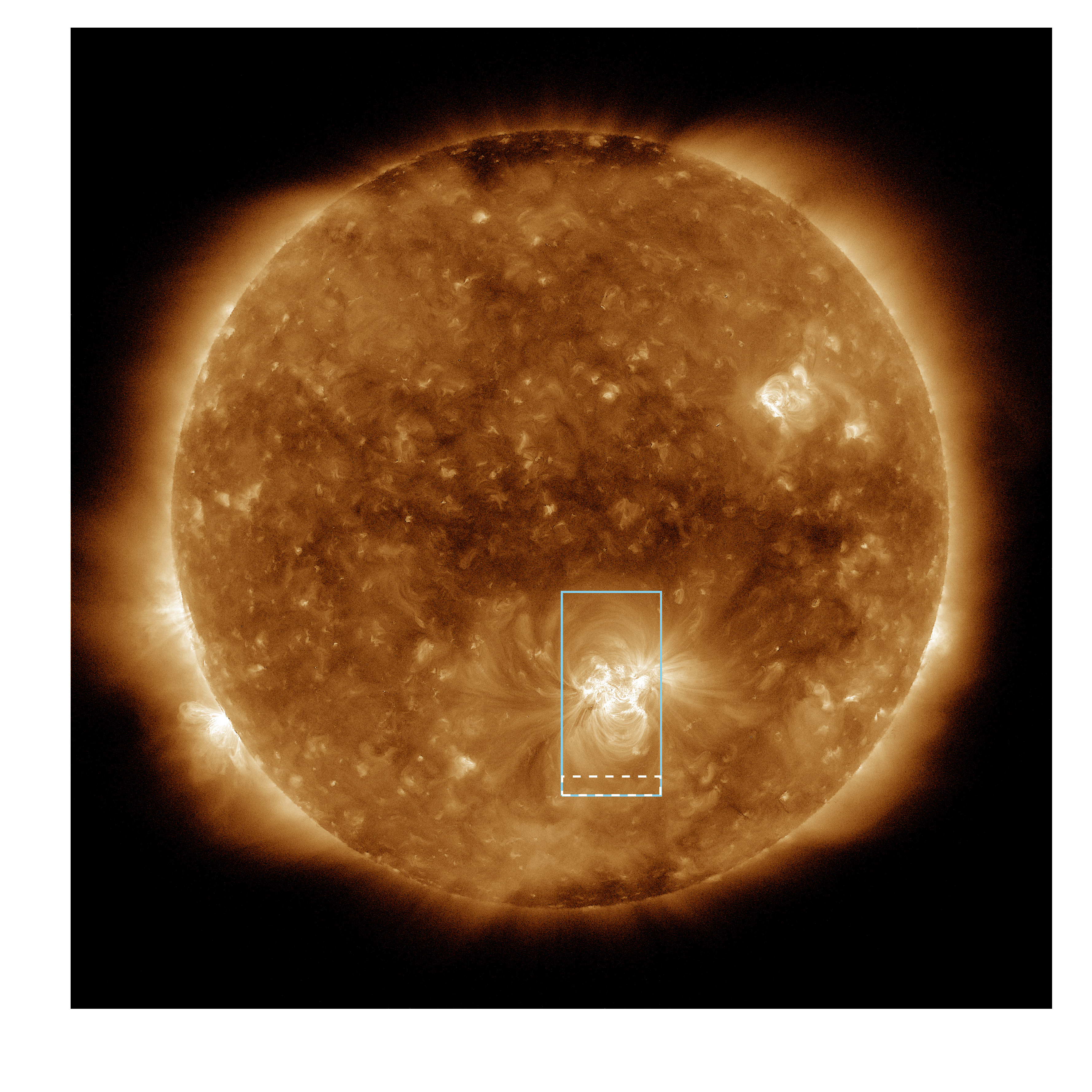}
\caption{ SDO/AIA 193\,\AA\, image of the solar corona showing AR 12781 on 2020, November 10, at 02:38:04\,UT. 
The sky blue box shows the field-of-view (FOV) of the EIS slit scan from 02:08:11--03:08:16\,UT. The white
dashed box shows the area used for velocity calibration (see text).
}
\label{fig1}
\end{figure}

For SPICE, we do not have sufficient temperature coverage from spectral lines of one species to perform the EM analysis for a
single element. We also do not have sufficient coverage to do the analysis for low-FIP elements only. We therefore follow a different
approach. We identify a sample of spectral lines from \ion{O}{2}--\ion{O}{6}, \ion{S}{4}--\ion{S}{5}, \ion{N}{3}--\ion{N}{4},
\ion{C}{3}, \ion{Ne}{6}, \ion{Ne}{8}, and \ion{Mg}{8}--\ion{Mg}{9} using the SUMER spectral atlas of on-disk features
\citep{Curdt2001}. We then compute the electron density using the \ion{Mg}{8} 772.31/782.34 diagnostic ratio, and calculate the $G(T,n)$ function for
all the spectral lines. The DEM is computed at this constant density again using the MCMC algorithm and assuming a fixed
set of elemental abundances. We then look for consistency between the observed and calculated intensities to determine which
set of abundances best represents the observations. 

Since most of the spectral lines we use ($\sim$ 80\%) are from high FIP elements, there is a possibility that changing the abundances for 
these elements could lead to better DEM convergence simply because they dominate the sample. This would make it appear that a FIP effect has been
detected, when in fact it is only a result of adopting a different set of abundances. To avoid any confusion as to the reasons why a particular abundance
set provides a better solution, in the case of SPICE, we adopt photospheric abundances for the high-FIP elements
even in the corona.
More precisely, only the low-FIP elements are enhanced in our coronal abundance dataset \citep[by the factor recommended by ][]{Schmelz2012}.

\begin{figure*}
\centering
\includegraphics[width=1.0\textwidth]{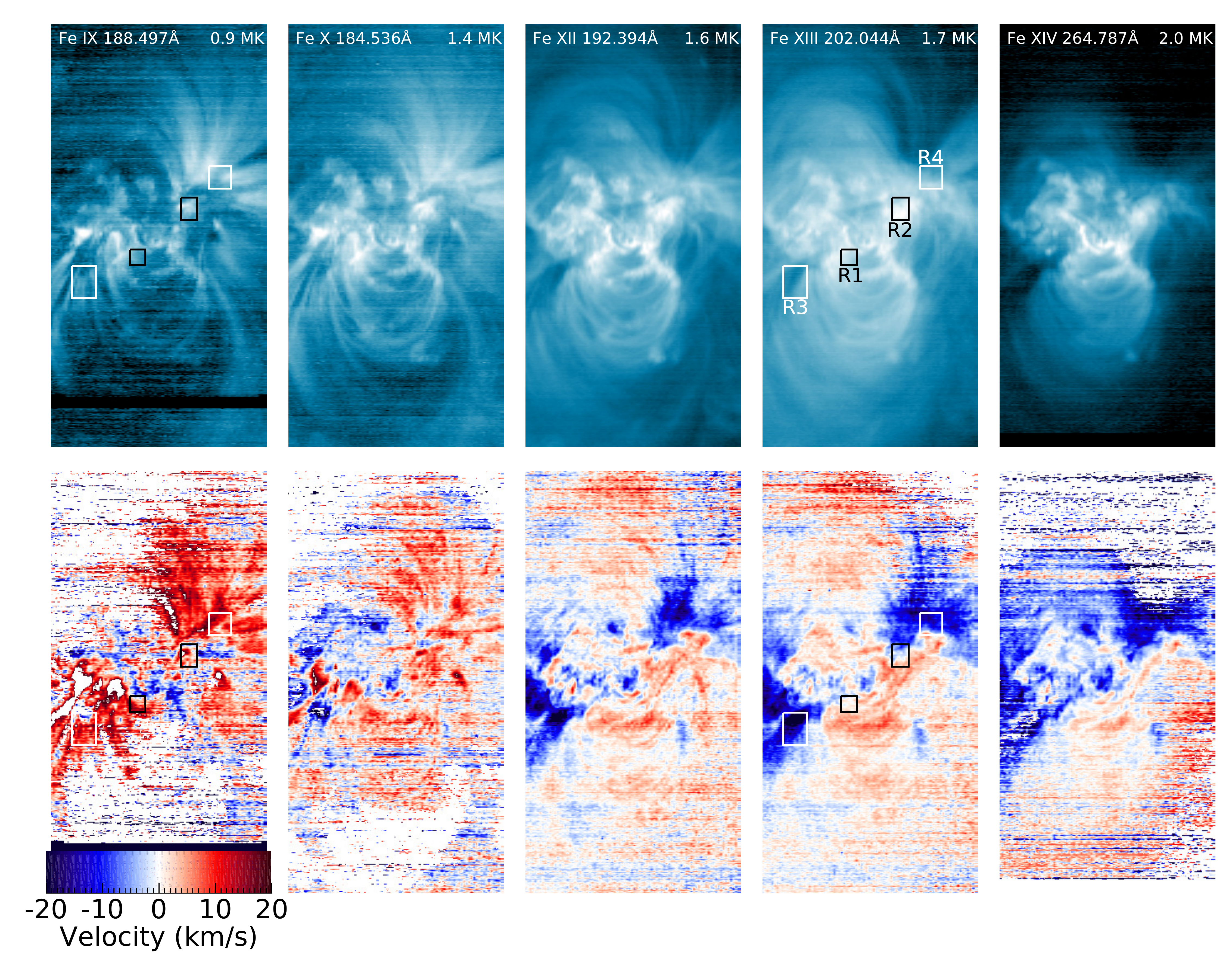}
\caption{ Intensity and velocity maps of AR 12781 constructed from the EIS scan from 02:08:11--03:08:16\,UT.
Upper row: Intensity maps covering a range of temperatures from 0.9--2.0\,MK using spectral lines of \ion{Fe}{9}--\ion{Fe}{14}.
Lower row: Velocity maps obtained from single Gaussian fits to the corresponding spectral lines.
Red/blue show plasma moving away/towards the observer. The white/black boxes labeled R1--R4 highlight the areas in the active region upflows/loop footpoints
selected for the elemental abundance analysis. 
}
\label{fig2}
\end{figure*}

\section{Results and discussion}
\label{discussion} 

AR 12781 appeared on the solar east limb on 2020, November 3. It was a bipolar region and produced numerous C-class flares
during its Earth-facing disk passage. According to the Hinode Flare Catalog \citep{Watanabe2012}, however, all but one of these flares 
(26/27), occurred prior to the EIS observations on November 10, so any impact on the compositional structure of AR 12781 between the EIS and SPICE observations
is likely to be minimal.
Figure \ref{fig1} shows the AR on a full disk AIA 193\,\AA\, image, with the EIS slit scan
FOV overlaid. Figure \ref{fig2}
shows examples of the EIS observations for a range of temperatures. Bright fan loops
can be seen to the east and west sides at lower temperatures ($\sim$0.9--1.4\,MK), with high lying loops in the central
structure ranging to higher temperatures ($\sim$0.9--2.0\,MK). Bright moss emission can be seen in the AR core. The cooler
fan loops are red-shifted, while characteristic blue-shifted upflows are seen at the AR edges; mixing with the fans.
These features are typical of EIS observations of large bipolar regions \citep{DelZanna2008,Brooks2011,Warren2011}.

We selected four regions in the AR for our composition analysis. R1 is a bright moss region at the base of high temperature 
loops and R2 is at the footpoint of a fan loop at the AR edge, while R3 and R4 were chosen to sit in the upflows that are seen as blue-shifted in the \ion{Fe}{13} 202.044\,\AA\, velocity map
of Figure \ref{fig2}. R2 and R4 are within the area that is later observed by SPICE.

\begin{figure*}
\centering
\includegraphics[width=0.45\textwidth]{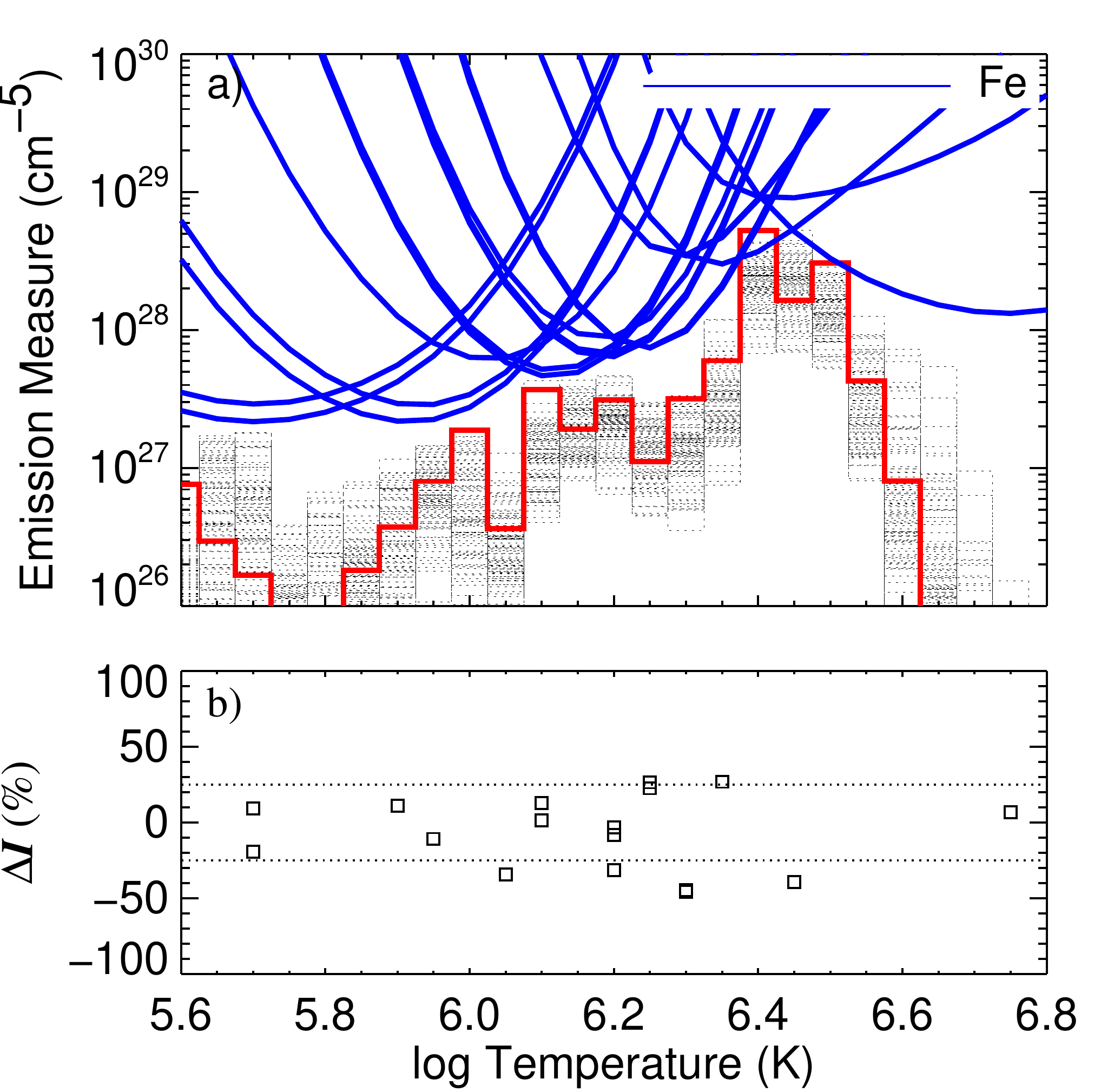}
\includegraphics[width=0.45\textwidth]{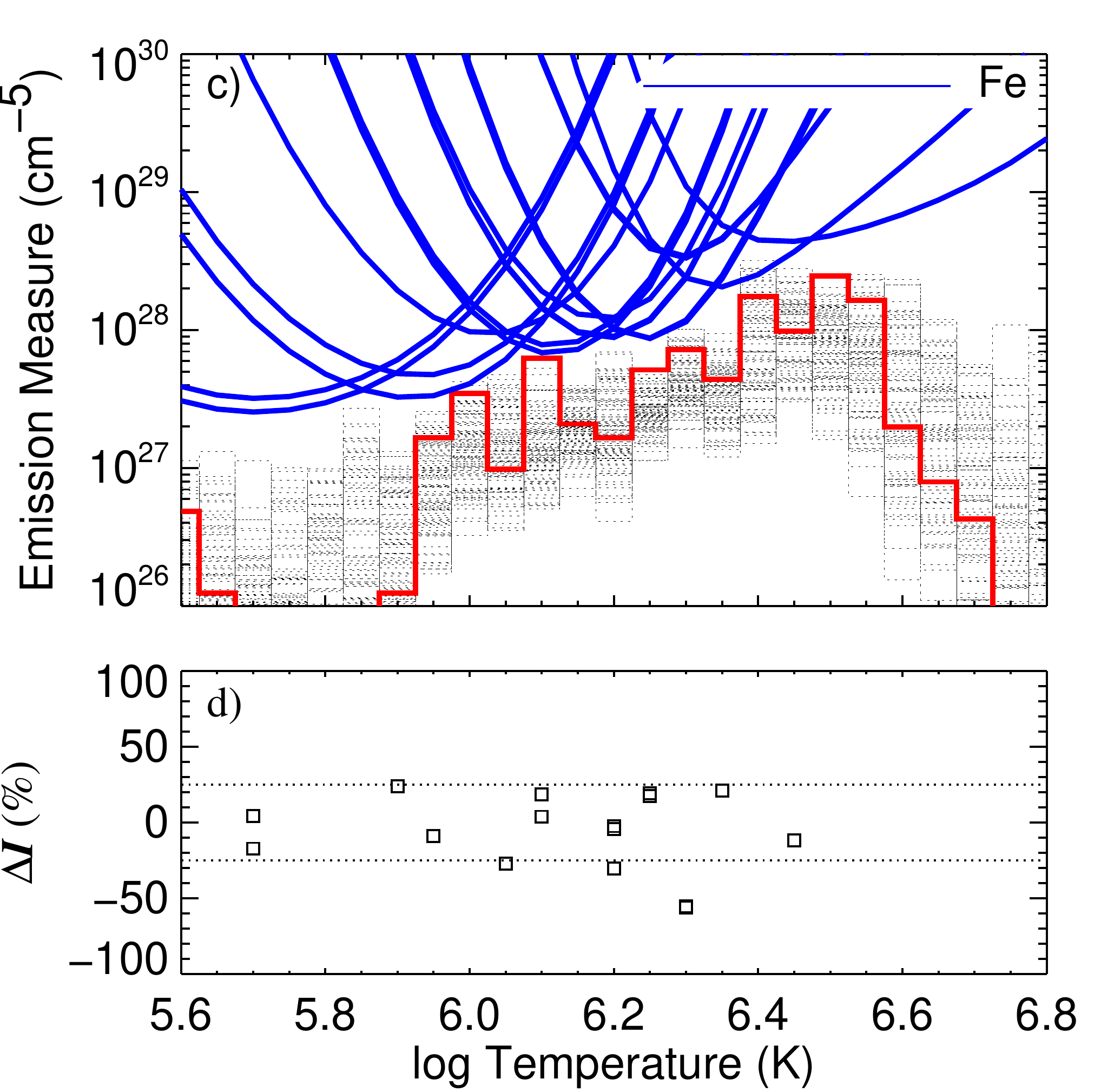}
\includegraphics[width=0.45\textwidth]{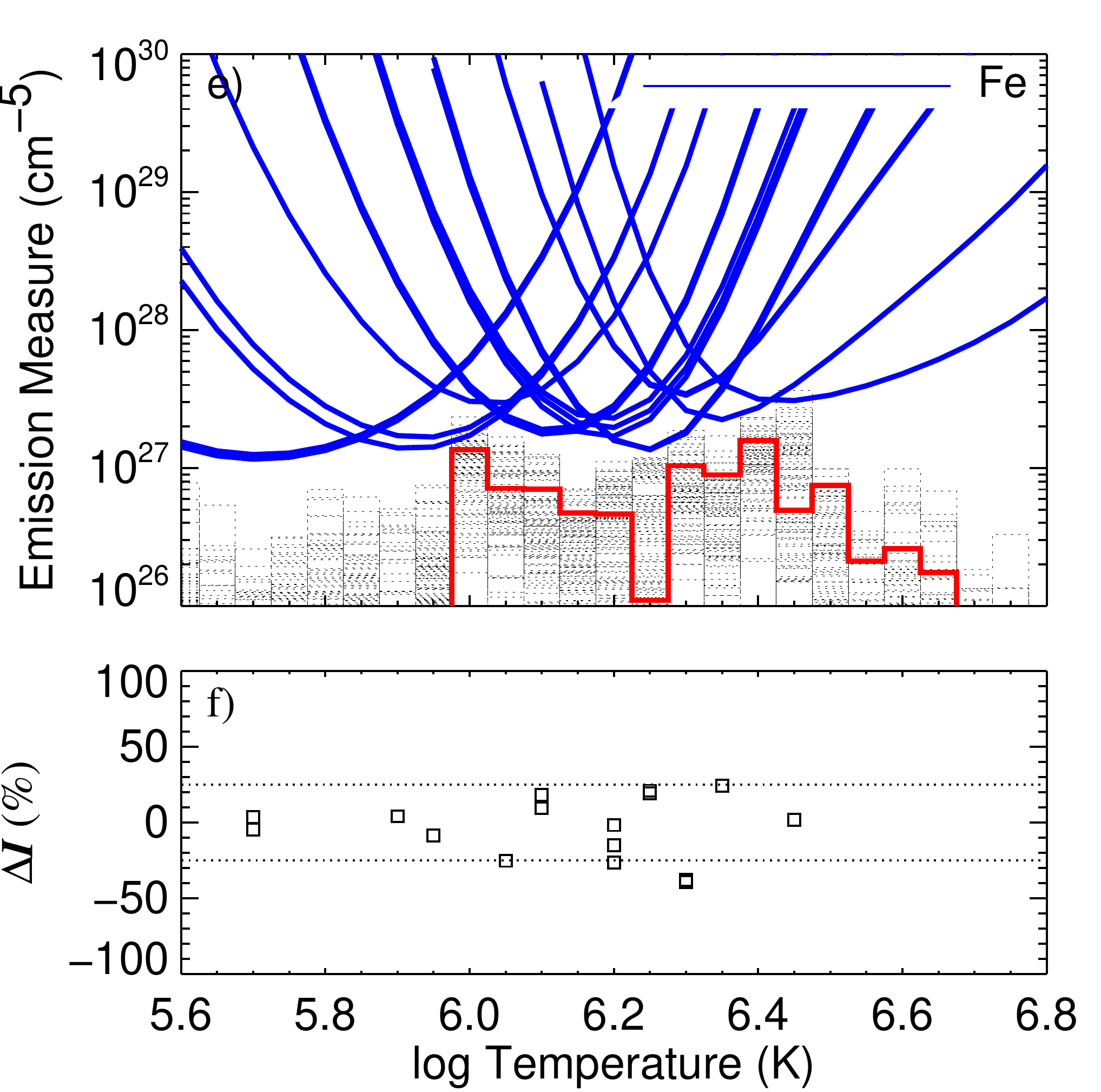}
\includegraphics[width=0.45\textwidth]{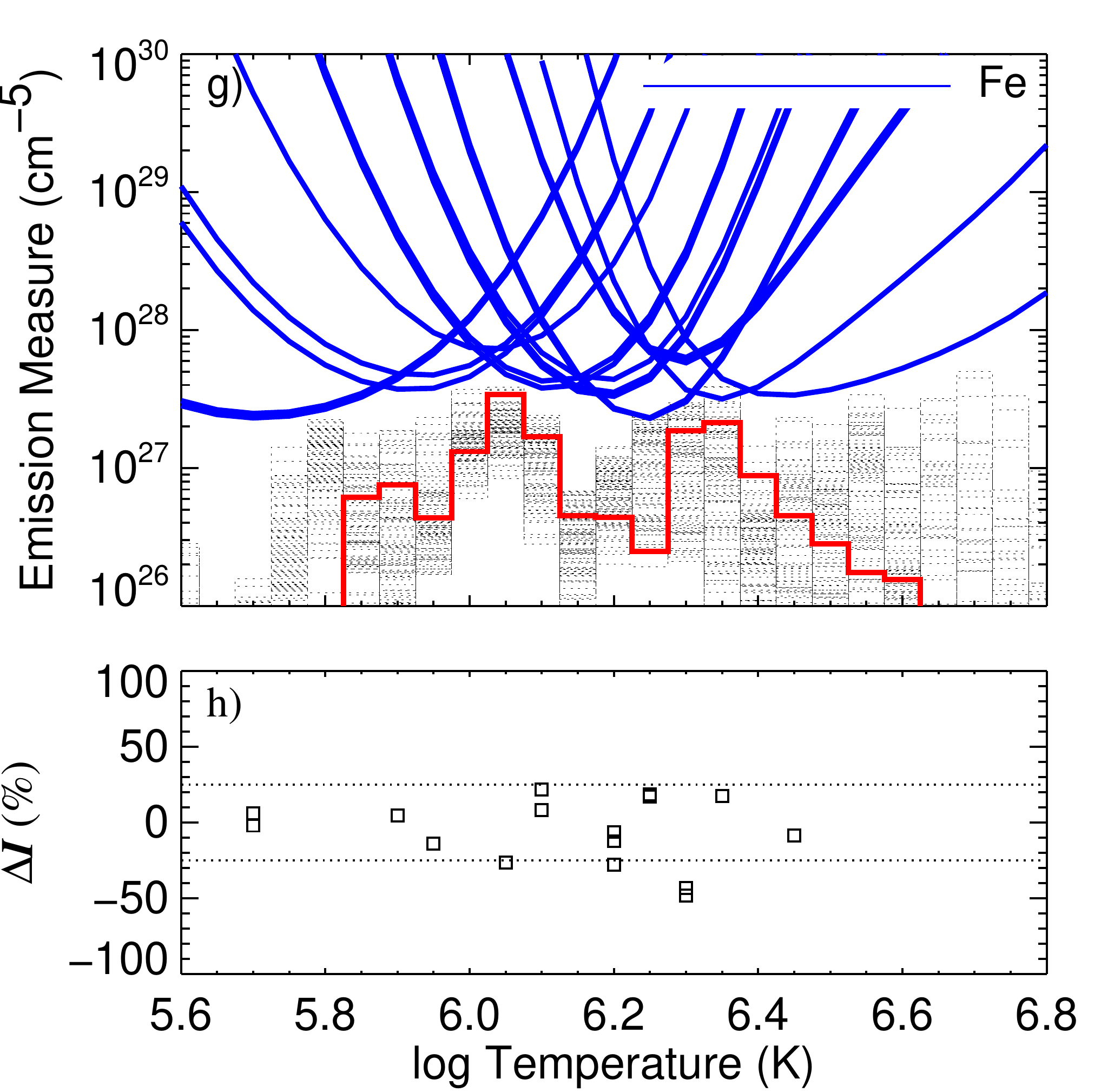}
\caption{ EIS emission measure (EM) analysis of the regions selected on 2020, November 10. 
The upper panels in each quadrant show the EM analysis. The red line shows the best-fit solution to a series of Monte Carlo simulations (shown as dotted
grey lines). The blue lines are emission measure loci curves for each of the spectral lines used in the analysis. These visualize the upper limits
to the emission measure for each line of each ion. The element for each line is also identified in the legend. 
The lower panels in each quadrant show the differences between the measured and calculated intensities. They are given as 
percentages of the observed intensities. The dotted lines indicate departures of 25\%. We show the results for region R1 in panels a \& b; R2 in panels c \& d;
R3 in panels e \& f; and R4 in panels g \& h.
}
\label{fig3}
\end{figure*}
We show our emission measure (EM) analysis of the four regions in Figure \ref{fig3}, together with a comparison of the observed
and EM calculated intensities. We also provide the observed and calculated intensities, and the percentage differences between them,
for all four regions in Table \ref{table1}. In most cases 
(80--90\%) the differences between the observed and 
calculated intensities are less
than $\sim$35\%; indicating that the best-fit MCMC solution is reproducing the observed intensities well. The \ion{Fe}{14} line pair 
(270.519\,\AA\, and 264.787\,\AA) seem to be consistently stronger than predicted in all four regions. This is clear from the EM
plots in Figure \ref{fig3}: the \ion{Fe}{14} loci curves that minimize at $\log$ T = 6.3, with T in units of K, sit above the loci curves for the
adjacent ions. The reasons for this discrepancy are unclear.

\begin{deluxetable*}{ccccccccccccc}
\tabletypesize{\footnotesize}
\tablecaption{EIS DEM analysis on 10-Nov-2020}
\tablehead{
\multicolumn{1}{c}{} &
\multicolumn{3}{c}{R1} &
\multicolumn{3}{c}{R2} &
\multicolumn{3}{c}{R3} &
\multicolumn{3}{c}{R4} \\
\multicolumn{1}{c}{ID} &
\multicolumn{1}{c}{I$_{obs}$} &
\multicolumn{1}{c}{I$_{calc}$} &
\multicolumn{1}{c}{$\Delta [\%]$} &
\multicolumn{1}{c}{I$_{obs}$} &
\multicolumn{1}{c}{I$_{calc}$} &
\multicolumn{1}{c}{$\Delta [\%]$} &
\multicolumn{1}{c}{I$_{obs}$} &
\multicolumn{1}{c}{I$_{calc}$} &
\multicolumn{1}{c}{$\Delta [\%]$} &
\multicolumn{1}{c}{I$_{obs}$} &
\multicolumn{1}{c}{I$_{calc}$} &
\multicolumn{1}{c}{$\Delta [\%]$} 
}
\startdata
Fe VIII 185.213 & 574.8$\pm$126.5 & 463.3 & -19.4 & 630.9$\pm$138.8 & 521.7 & -17.3 & 244.7$\pm$53.9 & 233.2 & -4.7 & 486.6$\pm$107.1 & 476.9 & -2.0\\
Fe VIII 186.601 & 297.9$\pm$65.6 & 325.2 & 9.2 & 350.0$\pm$77.0 & 365.3 & 4.4 & 160.1$\pm$35.2 & 166.0 & 3.7 & 318.4$\pm$70.1 & 337.7 & 6.1\\
Fe IX 188.497 & 108.3$\pm$23.9 & 120.1 & 10.9 & 159.9$\pm$35.2 & 198.0 & 23.8 & 67.5$\pm$14.9 & 70.3 & 4.1 & 179.9$\pm$39.6 & 188.0 & 4.5\\
Fe IX 197.862 & 61.9$\pm$13.6 & 55.2 & -10.9 & 104.9$\pm$23.1 & 95.6 & -8.8 & 39.9$\pm$8.8 & 36.5 & -8.6 & 112.2$\pm$24.7 & 96.5 & -14.0\\
Fe X 184.536 & 616.0$\pm$135.6 & 405.3 & -34.2 & 954.1$\pm$209.9 & 693.8 & -27.3 & 309.9$\pm$68.2 & 231.8 & -25.2 & 760.6$\pm$167.4 & 558.8 & -26.5\\
Fe XI 188.216 & 937.5$\pm$206.3 & 1058.9 & 13.0 & 1410.1$\pm$310.2 & 1671.5 & 18.5 & 387.7$\pm$85.3 & 458.3 & 18.2 & 832.4$\pm$183.1 & 1013.5 & 21.8\\
Fe XI 188.299 & 624.0$\pm$137.3 & 633.4 & 1.5 & 967.8$\pm$212.9 & 1004.4 & 3.8 & 255.0$\pm$56.1 & 279.2 & 9.5 & 570.9$\pm$125.6 & 617.2 & 8.1\\
Fe XII 195.119 & 1831.5$\pm$402.9 & 1680.3 & -8.3 & 2505.4$\pm$551.2 & 2401.0 & -4.2 & 599.3$\pm$131.8 & 509.5 & -15.0 & 1076.8$\pm$236.9 & 943.9 & -12.3\\
Fe XII 192.394 & 558.0$\pm$122.8 & 540.0 & -3.2 & 789.3$\pm$173.7 & 770.9 & -2.3 & 166.1$\pm$36.5 & 163.3 & -1.6 & 322.7$\pm$71.0 & 302.5 & -6.2\\
Fe XIII 202.044 & 946.7$\pm$208.3 & 1198.5 & 26.6 & 1263.9$\pm$278.1 & 1509.3 & 19.4 & 307.1$\pm$67.6 & 371.4 & 20.9 & 513.0$\pm$112.9 & 607.7 & 18.5\\
Fe XII 203.720 & 211.7$\pm$46.7 & 145.4 & -31.3 & 283.0$\pm$62.4 & 197.0 & -30.4 & 45.0$\pm$10.0 & 33.0 & -26.5 & 86.3$\pm$19.1 & 62.2 & -27.9\\
Fe XIII 203.826 & 1303.6$\pm$286.9 & 1598.6 & 22.6 & 1422.2$\pm$313.0 & 1669.2 & 17.4 & 152.8$\pm$33.7 & 182.4 & 19.4 & 260.8$\pm$57.5 & 305.6 & 17.2\\
Fe XIV 270.519 & 1728.5$\pm$380.3 & 955.5 & -44.7 & 1719.6$\pm$378.3 & 753.0 & -56.2 & 172.4$\pm$37.9 & 107.1 & -37.9 & 296.0$\pm$65.1 & 167.8 & -43.3\\
Fe XIV 264.787 & 3222.6$\pm$709.0 & 1745.7 & -45.8 & 2964.1$\pm$652.1 & 1327.6 & -55.2 & 279.7$\pm$61.5 & 170.3 & -39.1 & 515.7$\pm$113.5 & 267.3 & -48.2\\
Fe XV 284.160 & 15727.2$\pm$3460.0 & 19960.7 & 26.9 & 10778.2$\pm$2371.2 & 13035.2 & 20.9 & 1187.6$\pm$261.3 & 1475.2 & 24.2 & 1675.9$\pm$368.7 & 1968.1 & 17.4\\
Fe XVI 262.984 & 1860.5$\pm$409.3 & 1128.0 & -39.4 & 898.7$\pm$197.7 & 791.8 & -11.9 & 62.9$\pm$13.8 & 64.0 & 1.7 & 68.9$\pm$15.2 & 63.0 & -8.5\\
Fe XVII 254.87 & 19.7$\pm$4.4 & 21.1 & 6.8 &  &  &  &  &  &  &  &  &   
\enddata
\tablenotetext{}{EIS line intensities are in units of erg cm$^{-2}$ s$^{-1}$ sr$^{-1}$.}
\label{table1}
\end{deluxetable*}

\begin{figure}[h]
\centering
\includegraphics[width=0.5\textwidth]{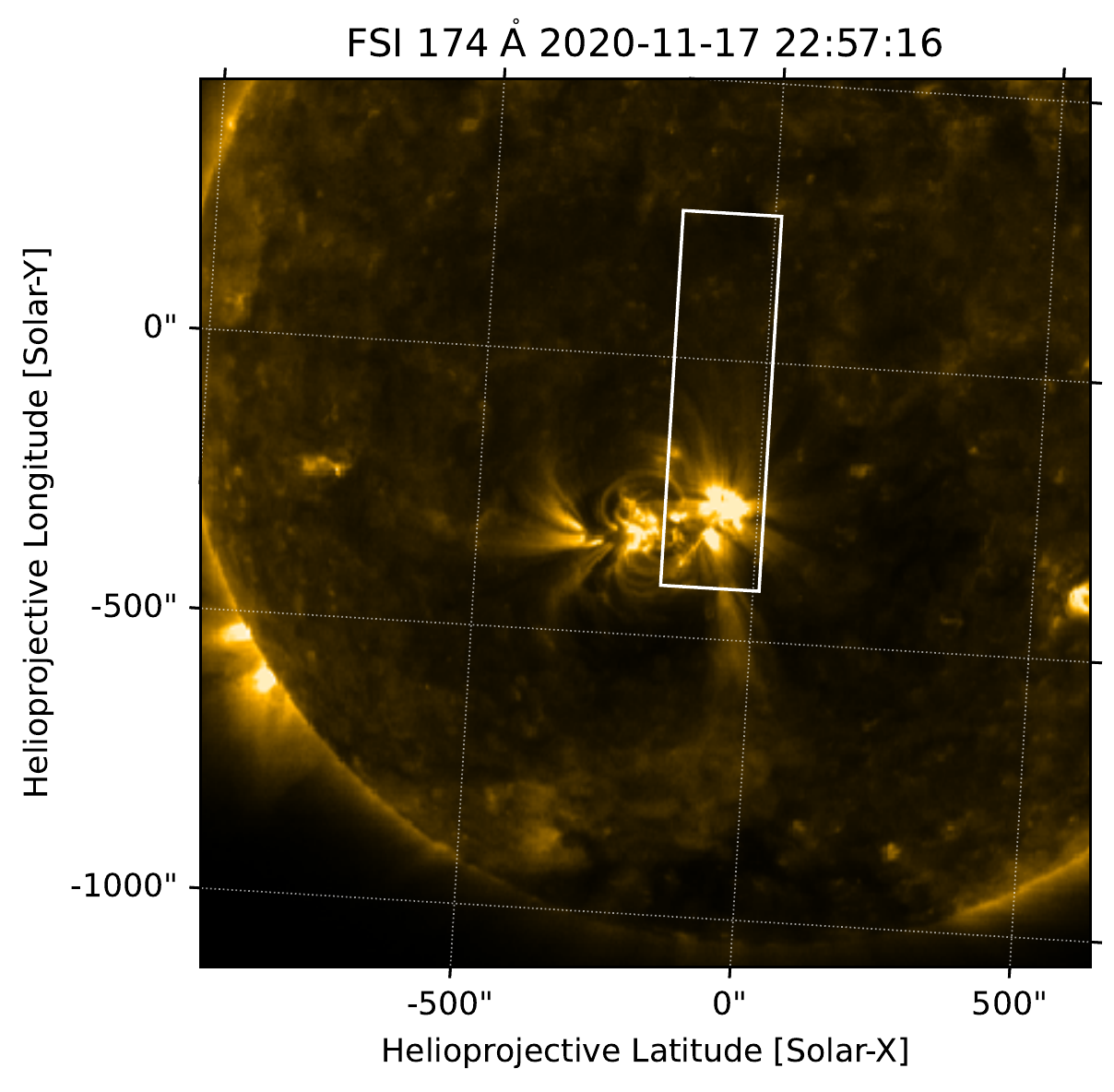}
\caption{ Solar Orbiter EUI/FSI 174\,\AA\, image of the solar corona showing AR 12781 on 2020, November 17, at 22:57:16\,UT. 
The white box shows the FOV of the SPICE slit scan from 22:28:26--23:19:40\,UT. 
}
\label{fig4}
\end{figure}

\begin{figure*}
\centering
\includegraphics[width=1.0\textwidth]{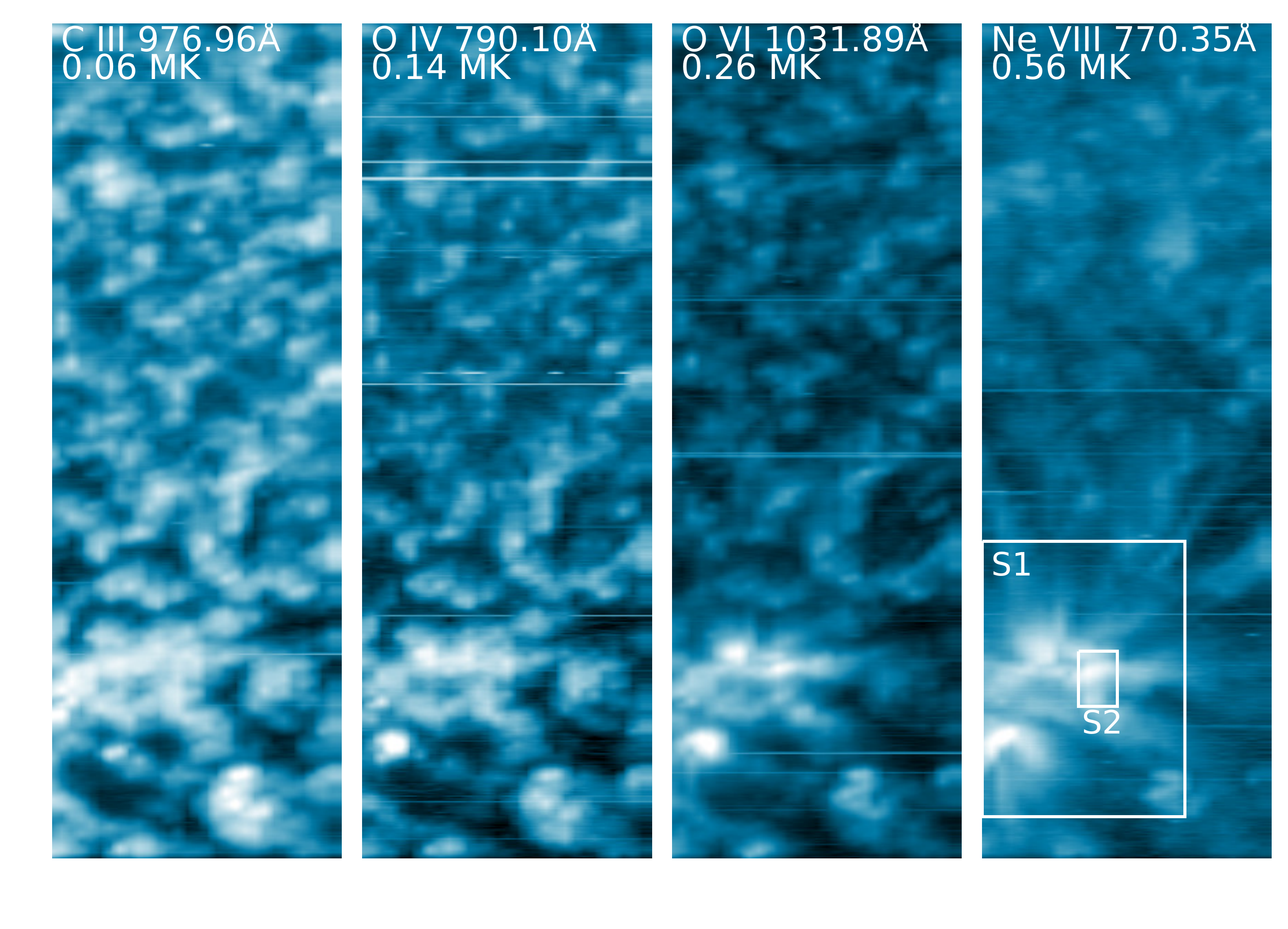}
\caption{ Intensity maps of AR 12781 constructed from the SPICE scan from 22:28:26--23:19:40\,UT.
The panels show intensity maps covering a range of temperatures from 0.06--0.56\,MK using spectral lines of \ion{C}{3}, \ion{O}{4}, \ion{O}{6}, and \ion{Ne}{8}.
The white boxes labeled S1 and S2 highlight the areas in the active region/loop footpoint
selected for the elemental abundance analysis. 
}
\label{fig5}
\end{figure*}

\begin{figure*}
\centering
\includegraphics[width=0.45\textwidth]{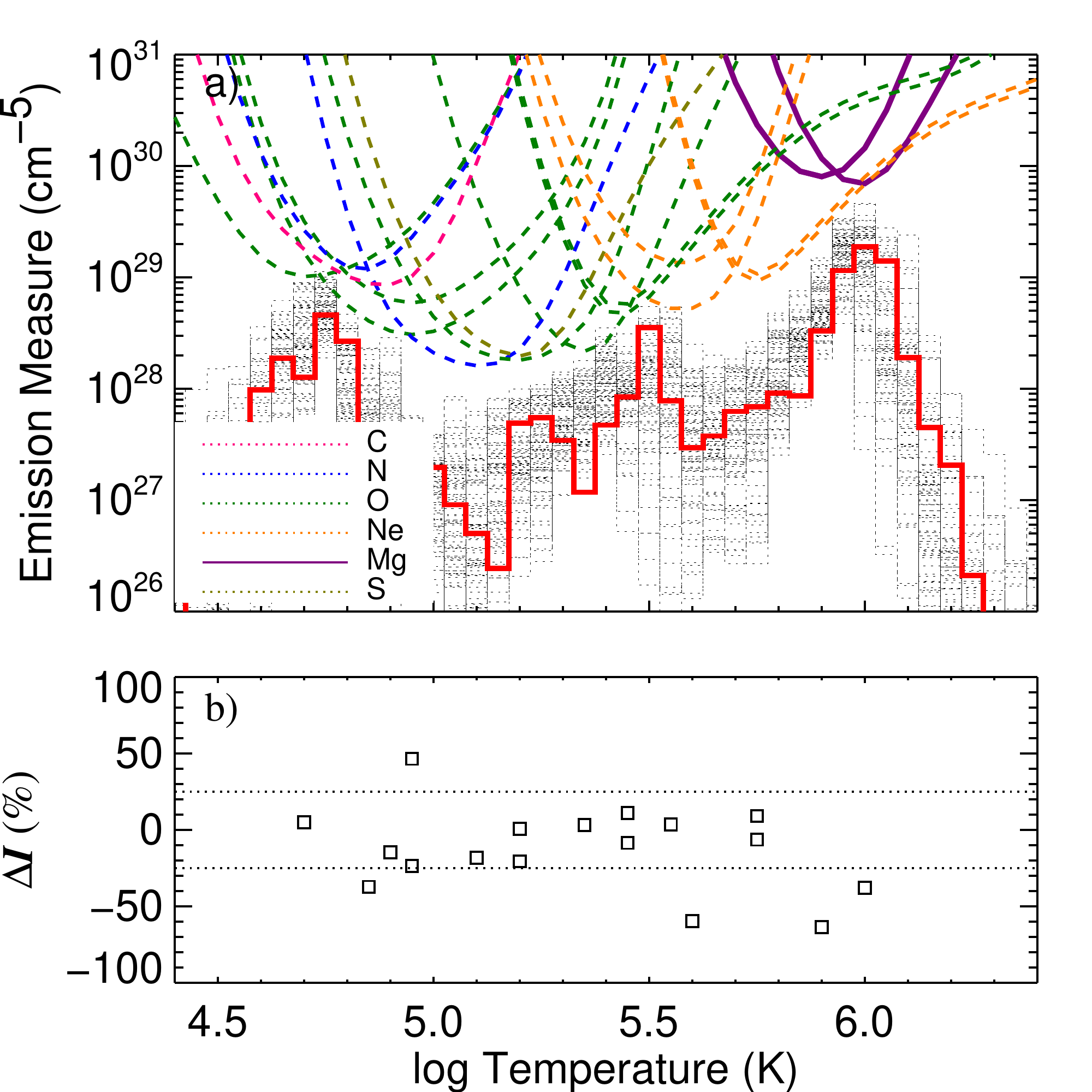}
\includegraphics[width=0.45\textwidth]{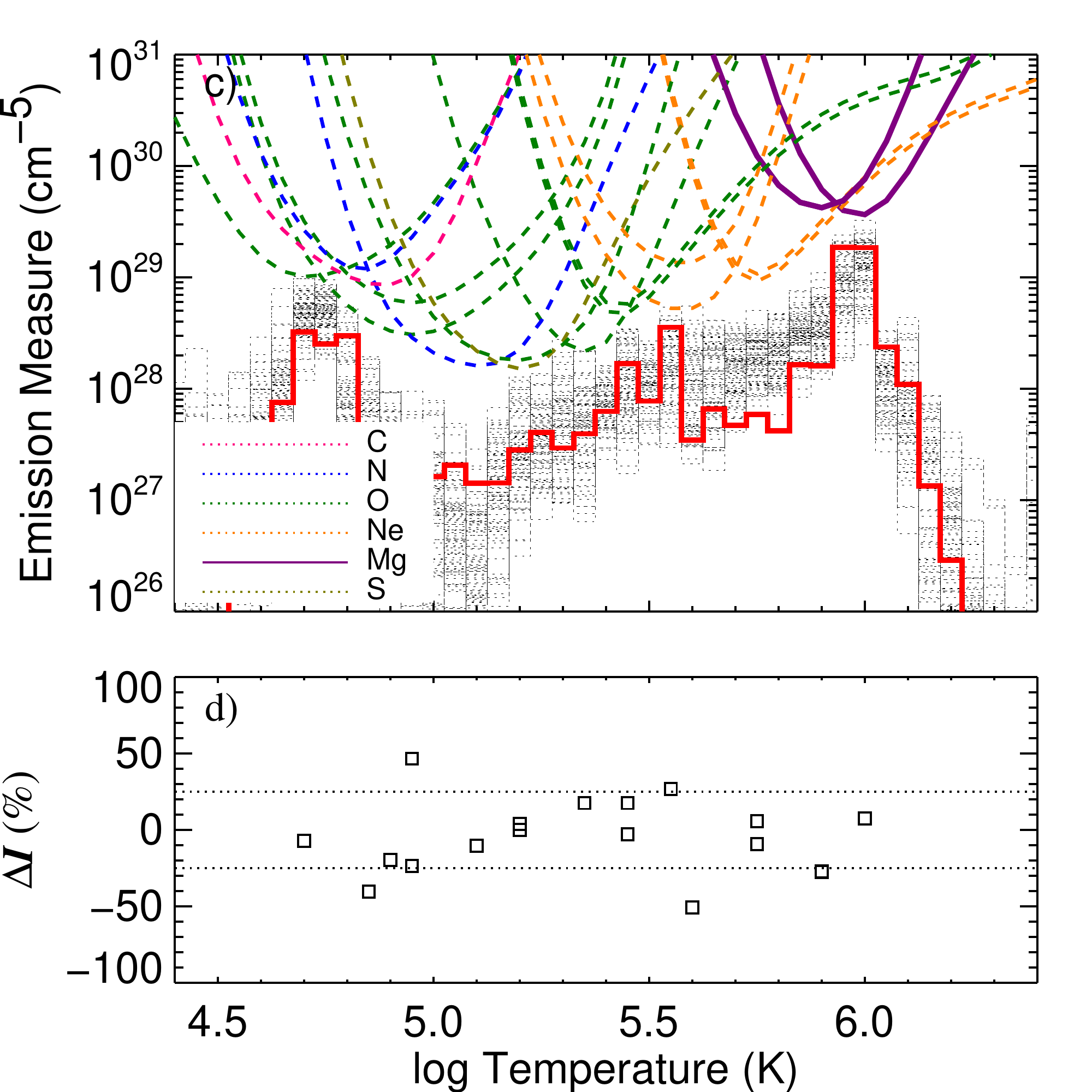}
\includegraphics[width=0.45\textwidth]{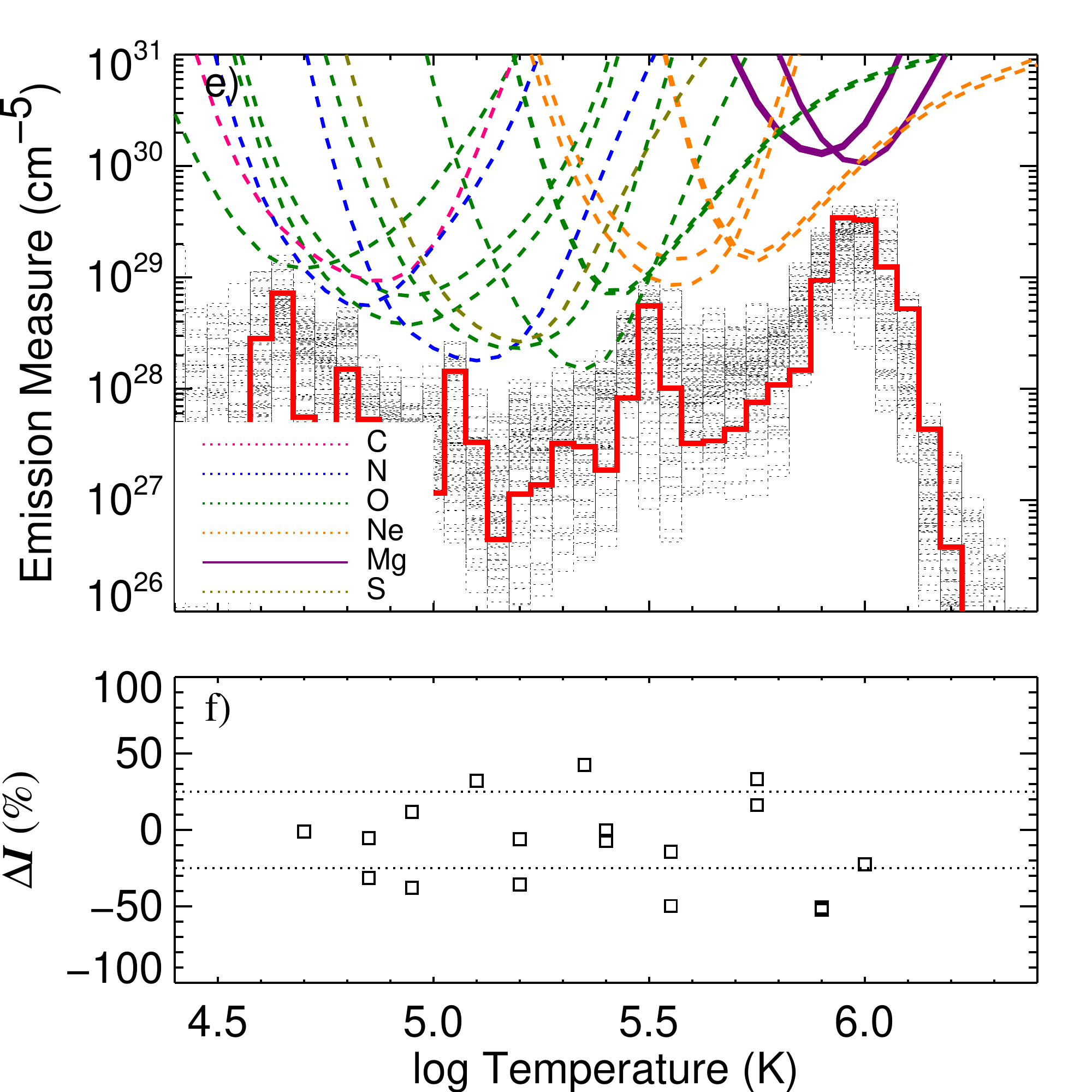}
\includegraphics[width=0.45\textwidth]{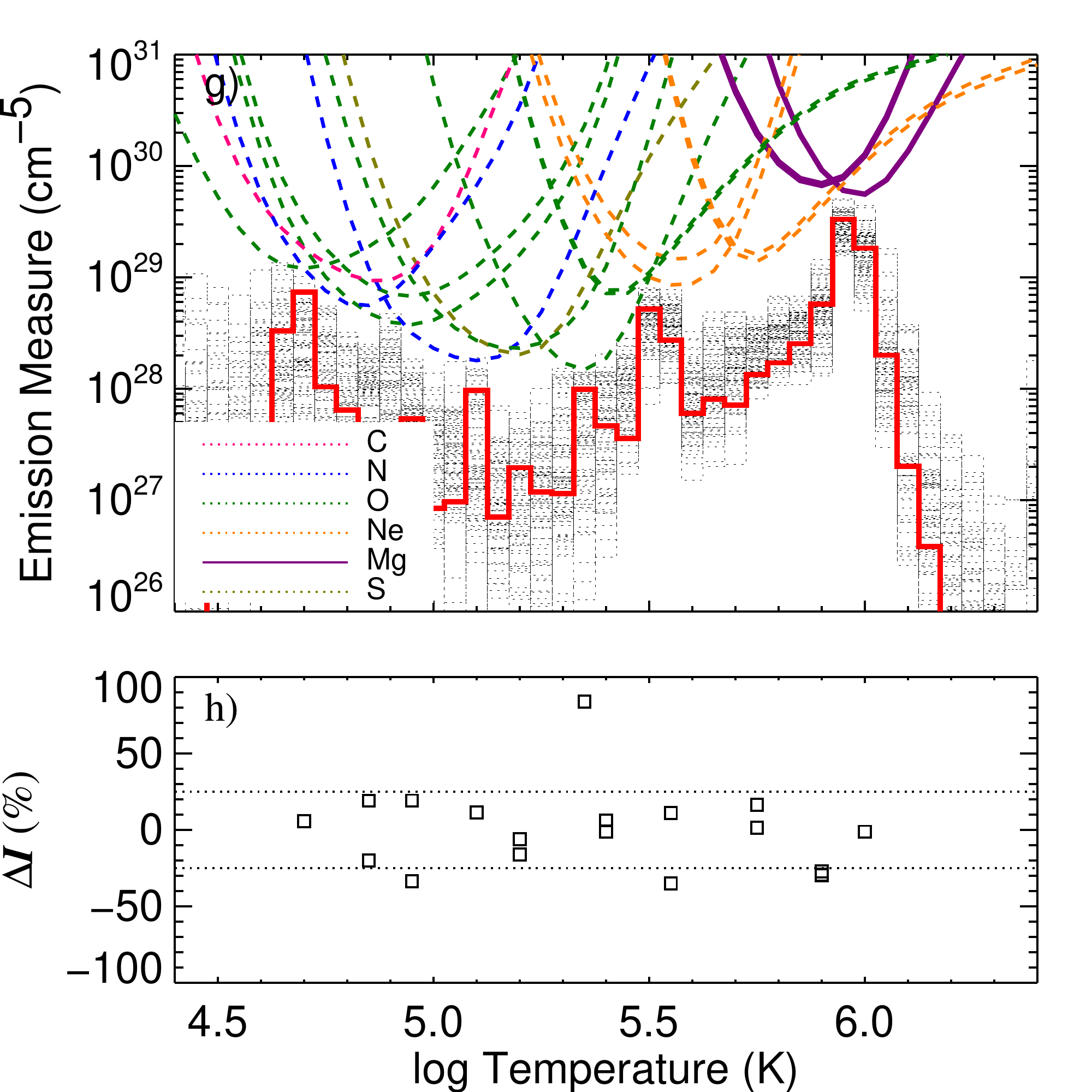}
\caption{ SPICE emission measure analysis of the regions selected on 2020, November 17. 
The upper panels in each quadrant show the EM analysis. The red line shows the best-fit solution to a series of Monte Carlo simulations (shown as dotted
grey lines). The colored lines are emission measure loci curves for each of the spectral lines used in the analysis. These visualize the upper limits
to the emission measure. The element for each line is identified by color in the legend. The solid loci curves indicate lines from low-FIP elements and the dashed loci curves
indicate lines from high-FIP elements.
The lower panels in each quadrant show the differences between the measured and calculated intensities. They are given as 
percentages of the observed intensities. The dotted lines indicate departures of 25\%. We show the photospheric results for region S1 in panels a \& b; the coronal results
for S1 in panels c \& d;
the photospheric results for S2 in panels e \& f; and the coronal results for S2 in panels g \& h.
}
\label{fig6}
\end{figure*}

\begin{figure}[h]
\centering
\includegraphics[width=0.5\textwidth]{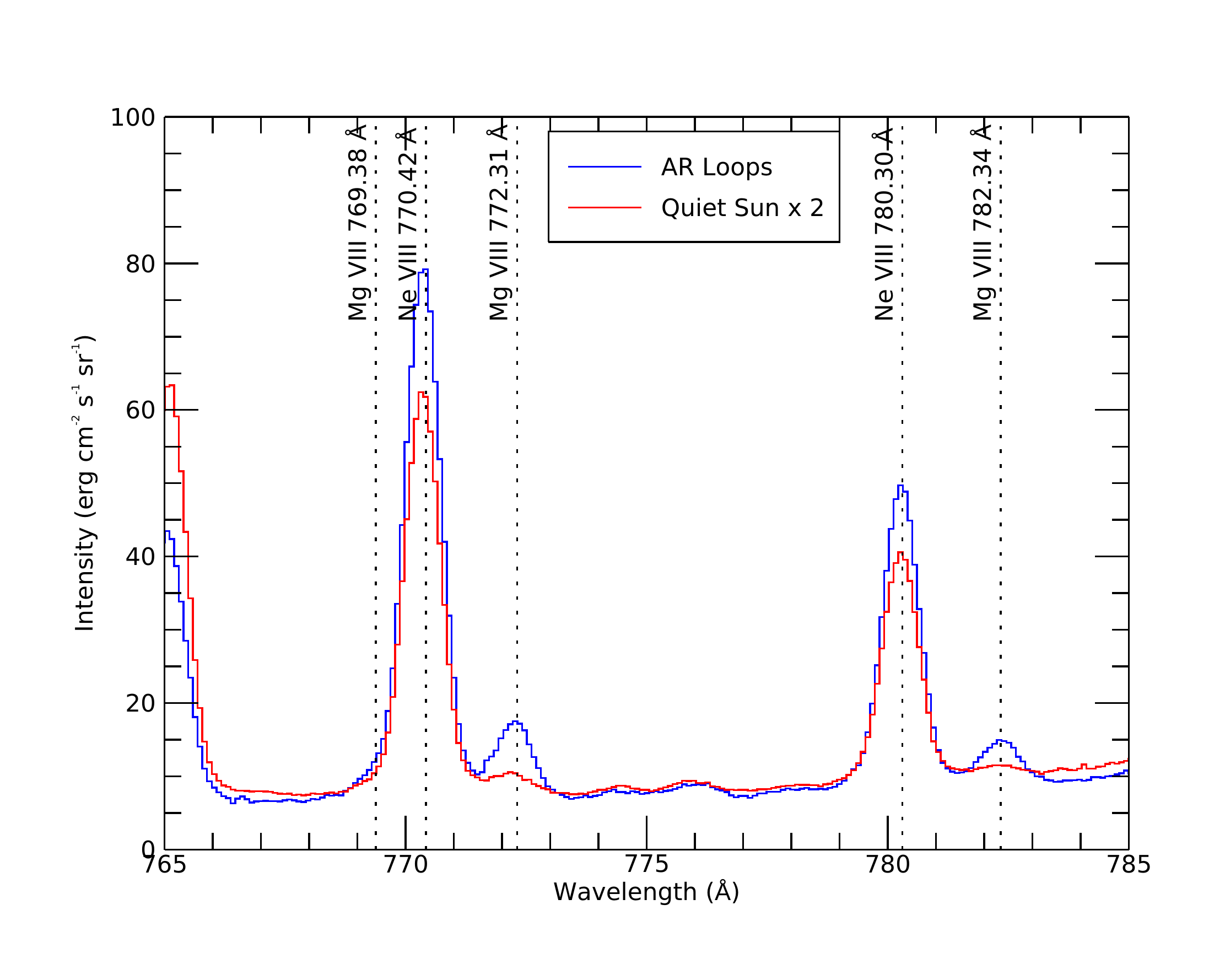}
\caption{ Comparison of loop footpoint (blue; region S2) and quiet Sun (red) SPICE spectra for the 765--785\,\AA\, wavelength range where critical low FIP \ion{Mg}{8} 
and high FIP \ion{Ne}{8} spectral lines fall. We scaled the quiet Sun spectrum by a factor of 2 in order to show the relative strengths of the 
\ion{Mg}{8} lines. They are stronger when the low FIP elements are enhanced in the corona. The line identifications
are given in the vertical legends.
}
\label{fig7}
\end{figure}

\begin{deluxetable*}{ccccccccccc}
\tabletypesize{\small}
\tablecaption{SPICE DEM analysis on 17-Nov-2020}
\tablehead{
\multicolumn{2}{c}{} &
\multicolumn{2}{c}{S1 - Photospheric} &
\multicolumn{2}{c}{S1 - Coronal} &
\multicolumn{1}{c}{} &
\multicolumn{2}{c}{S2 - Photospheric} &
\multicolumn{2}{c}{S2 - Coronal} \\
\multicolumn{1}{c}{ID} &
\multicolumn{1}{c}{I$_{obs}$} &
\multicolumn{1}{c}{I$_{calc}$} &
\multicolumn{1}{c}{$\Delta [\%]$} &
\multicolumn{1}{c}{I$_{calc}$} &
\multicolumn{1}{c}{$\Delta [\%]$} &
\multicolumn{1}{c}{I$_{obs}$} &
\multicolumn{1}{c}{I$_{calc}$} &
\multicolumn{1}{c}{$\Delta [\%]$} &
\multicolumn{1}{c}{I$_{calc}$} &
\multicolumn{1}{c}{$\Delta [\%]$} 
}
\startdata
O III 702.61 & 20.2$\pm$6.1 & 15.4 & -23.6 & 15.4 & -23.6 & 22.3$\pm$6.6 & 13.9 & -37.8 & 14.8 & -33.6\\
O III 703.87 & 19.2$\pm$5.7 & 28.2 & 46.6 & 28.2 & 46.6 & 22.7$\pm$6.5 & 25.3 & 11.7 & 27.0 & 19.1\\
Mg IX 706.02 & 12.5$\pm$3.5 & 7.7 & -37.9 & 13.4 & 7.5 & 19.1$\pm$5.1 & 14.8 & -22.5 & 18.8 & -1.2\\
O II 718.49 & 7.2$\pm$2.2 & 7.5 & 5.0 & 6.6 & -7.2 & 7.8$\pm$2.4 & 7.8 & -1.0 & 8.3 & 5.7\\
O V 760.43 & 11.2$\pm$5.0 & 11.6 & 3.3 & 13.2 & 17.7 & 2.2$\pm$1.9 & 3.1 & 42.6 & 4.0 & 83.9\\
N IV 765.15 & 30.1$\pm$8.0 & 24.7 & -18.1 & 27.0 & -10.3 & 32.8$\pm$8.6 & 43.4 & 32.2 & 36.6 & 11.5\\
Mg VIII 769.38 &  &  &  &  &  & 2.2$\pm$1.9 & 1.1 & -50.8 & 1.6 & -27.1\\
Ne VIII 770.42 & 42.8$\pm$11.2 & 46.8 & 9.1 & 45.3 & 5.8 & 67.2$\pm$17.2 & 89.5 & 33.3 & 78.2 & 16.4\\
Mg VIII 772.31 & 5.7$\pm$2.4 & 2.1 & -63.6 & 4.2 & -27.3 & 10.4$\pm$3.3 & 4.9 & -52.2 & 7.3 & -29.3\\
Ne VIII 780.30 & 24.7$\pm$6.6 & 23.1 & -6.3 & 22.4 & -9.2 & 38.1$\pm$9.9 & 44.3 & 16.3 & 38.7 & 1.5\\
Mg VIII 782.34 & 4.5$\pm$2.4 & 1.6 & -63.7 & 3.2 & -27.4 & 6.8$\pm$2.7 & 3.2 & -52.3 & 4.8 & -29.5\\
S V 786.47 & 16.2$\pm$4.8 & 12.9 & -20.5 & 16.2 & 0.0 & 21.7$\pm$6.2 & 14.0 & -35.6 & 18.2 & -16.0\\
O IV 787.72 & 30.5$\pm$8.3 & 30.7 & 0.7 & 31.7 & 4.1 & 38.6$\pm$10.3 & 36.3 & -6.0 & 36.3 & -6.1\\
C III 977.03 & 415.2$\pm$104.0 & 354.3 & -14.7 & 334.1 & -19.5 & 430.7$\pm$107.9 & 295.2 & -31.5 & 344.2 & -20.1\\
N III 991.55 & 25.8$\pm$7.3 & 16.1 & -37.3 & 15.3 & -40.4 & 12.3$\pm$4.8 & 11.6 & -5.4 & 14.6 & 19.1\\
Ne VI 999.27 & 5.9$\pm$2.2 & 2.4 & -59.8 & 2.9 & -50.8 & 6.6$\pm$2.3 & 3.3 & -49.8 & 4.3 & -34.9\\
Ne VI 1005.79 & 1.4$\pm$1.2 & 1.4 & 3.6 & 1.8 & 26.8 & 2.3$\pm$1.4 & 1.9 & -14.3 & 2.5 & 11.0\\
O VI 1031.93 & 175.6$\pm$44.7 & 194.8 & 10.9 & 206.6 & 17.7 & 273.7$\pm$69.0 & 272.6 & -0.4 & 290.9 & 6.3\\
O VI 1037.64 & 105.7$\pm$27.9 & 96.8 & -8.4 & 102.7 & -2.9 & 145.9$\pm$38.5 & 135.5 & -7.1 & 144.5 & -0.9  
\enddata
\tablenotetext{}{SPICE line intensities have been converted to units of erg cm$^{-2}$ s$^{-1}$ sr$^{-1}$.}
\label{table2}
\end{deluxetable*}
The loop footpoint regions (R1 and R2) show EM distributions that are quite strongly peaked around 2.5--3.2\,MK. This is quite typical
of the EM distributions found in AR cores \citep{Warren2012}, albeit a slightly lower temperature. The densities in these regions
were measured at $\log$ n = 9.1 and 9.0, with n in units of cm$^{-3}$, and the FIP bias measured using the Si/S ratio is 3.1 and 2.5 in R1 and R2, respectively.
The EM distributions for the upflow regions (R3 and R4) are somewhat flatter. Based on the velocity maps in Figure \ref{fig2} it is likely
that the emission from lines below 1\,MK\, do not come from the upflows. The fan loops are red-shifted and there is contamination
from foreground and background emission that complicates the interpretation. Furthermore, the high temperature \ion{Fe}{17} lines are too weak to measure in these quieter regions (also in R2). Nevertheless the density in both these
regions is $\log$ n = 8.6, and the FIP bias measured using the Si/S ratio is 1.6 and 1.2 in R3 and R4, respectively. 

The FIP related enhancement in upflow region R3 is relatively weak, though probably
above the uncertainty level so that it is likely real. In region R4 the FIP bias is consistent with photospheric abundances, or the fractionation of S. Similarly
variable results have been found for the upflows before \citep{Brooks2020}. 

AR 12781 rotated out of the Earth-facing view and arrived in the SPICE FOV on 2020 November 17. 
Figure \ref{fig4} shows the AR on an Extreme Ultraviolet Imager \citep[EUI,][]{Rochus2020} Full Sun Imager (FSI) 174\,\AA\, image, with the SPICE slit scan
FOV overlaid. Figure \ref{fig5}
shows examples of the SPICE observations for a range of temperatures below those accessed by EIS observations. The bright chromospheric
network can be seen at 0.06--0.26\,MK. It appears that the SPICE FOV mainly captures the western fan loop region
observed by EIS and shown in Figure \ref{fig2}. The \ion{Ne}{8} 770.35\,\AA\, intensity image shows the bright
moss emission at the base of the fan loops. Currently, there is an ongoing effort to understand the impact of systematic biases
in SPICE Doppler velocity measurements linked to peculiarities of the telescope and spectrometer; based on the EIS \ion{Fe}{9} 188.497\,\AA\, velocity map, however, these structures are likely to be red-shifted. 

The \ion{Ne}{8} 770.35\,\AA\, line is formed around 0.56\,MK. Unfortunately, this temperature
is too low to see the upflows, which are usually
seen above $\sim$1\,MK in EIS observations, and 
construction of SPICE velocity maps at these higher temperatures (e.g. using \ion{Mg}{9}) is difficult. 
We focused on two regions in the SPICE FOV. S1 encompasses most of the bright fan region and was chosen to make
sure a strong signal was obtained. S2 is at the base of the fan loops and was chosen for comparison
with the EIS observations. The fan loops on the western side of the AR do not appear markedly different from when they were 
observed by EIS, and region S2 is approximately the same area as R4. We were not able to obtain reliable results for region R2.

We show our EM analysis of the two regions in Figure \ref{fig6}, together with a comparison of the observed
and EM calculated intensities. We also provide the observed and calculated intensities, and the percentage differences between them,
for both regions in Table \ref{table2}. Note that we provide the SPICE intensities 
in the original SI units but that we convert them to cgs units
for consistency with the EIS intensities when calculating the EM.
Also, although we identified and attempted to fit about 26 lines in the SPICE spectra, several lines were either too weak to be useful or had errors larger than the measured intensities. We removed these
from our EM analysis. 
Several lines from \ion{O}{4} and \ion{O}{5} were excluded in this way, but more importantly, several high-FIP element lines were also affected. \ion{S}{4} 748.40\,\AA\, and \ion{S}{4} 750.20\,\AA\, may be useful
in other studies of elemental abundances in the lower transition region, but are not critical for comparisons with EIS coronal measurements. The \ion{Mg}{9} 749.54\,\AA\, line, however, could have been useful for 
our analysis but was very weak in our spectra. \ion{Mg}{8} 769.38\,\AA\, was also affected in region S1 (see Table \ref{table2}).

As discussed, we do not have a good abundance diagnostic ratio from two lines formed in the
same temperature range, so here we assess how well the line intensities are reproduced depending on whether we adopt photospheric
or coronal abundances for the EM analysis (with special emphasis on the \ion{Mg}{8} lines). 
Figure \ref{fig6} and Table \ref{table2} show the results for both these 
calculations for both regions.

The purpose of the MCMC algorithm is to minimize the differences between the observed and calculated intensities, so it is
no surprise that a reasonable solution is obtained for all the examples. When adopting photospheric abundances,
$\sim$65\% of the line intensities are reproduced to within
$\sim$35\% for both regions S1 and S2. There is a clear improvement, however, when using coronal abundances: 
85--95\% of the line intensities are reproduced. There is some scatter in the behavior for the lines from high-FIP elements: some of them are
better reproduced, while others become worse. For plasma composition studies, the behavior of the lines from high-FIP elements is the main interest. The only 
S line in the analysis is \ion{S}{5} 786.47\,\AA. It is reproduced in both regions with photospheric abundances, but the agreement is improved with coronal abundances. 
For this analysis, S is treated as a high-FIP element so the assumed abundance is the same in both datasets. As we discussed in section \ref{adam}, S often
shows unusual behavior and can act like a low-FIP element. Note, however, that if we enhance the S abundance by the factor given by \cite{Schmelz2012} then the agreement between
calculated and observed intensity worsens. This suggests that treating S as a high-FIP element is valid for this dataset.
However, the critical point is that the \ion{Mg}{8}
lines show a clear difference. \ion{Mg}{8} 772.31\,\AA, and \ion{Mg}{8} 782.34\,\AA\, 
are too weak by factors of 2.1--2.8 for both regions S1 and S2 when photospheric abundances are used, and \ion{Mg}{8} 769.38\,\AA\, is also too weak
by a factor of 2 in region S2. In contrast, 
all three lines are reproduced when coronal abundances are assumed. This can also be seen in the EM plots, where
the \ion{Mg}{8} EM loci curves sit above the general trend when photospheric abundances are used (panels a and e), but
are more consistent with the other curves when coronal abundances are assumed (panels c and g). 
We show example spectra to illustrate the behavior of the \ion{Mg}{8} lines compared to the \ion{Ne}{8} lines in Figure \ref{fig7}.
We clearly see that the \ion{Mg}{8} lines are strong in footpoint region S2 relative to a quiet Sun control spectrum (dashed box
region in Figure \ref{fig4}).
Significantly, the calculated intensities for all the \ion{Mg}{8} and \ion{Mg}{9} lines in S1, and the \ion{Mg}{8} 772.31\,\AA, and \ion{Mg}{8} 782.34\,\AA\, lines in S2,
are outside the error bars on the measured intensities when photospheric abundances are assumed. None of the calculated intensities for these lines are outside the 
error bars when coronal abundances are adopted.
Effectively there is no temperature structure that can explain the Mg and Ne intensities simultaneously with photospheric abundances.

The EM distributions
themselves show expected characteristics: the magnitude of the EM is higher in the transition region and low corona than measured by EIS, and the curves fall to a minimum in the $\log$ T = 5.0--5.3 range. There is some complex structure in the EM distributions that can be understood as arising from the MCMC algorithm reducing the differences between the observed and calculated intensities before the curves then rise
with increasing temperature until the upper boundary defined by the \ion{Mg}{9} lines is reached. Recall that in section \ref{adam} we 
discussed how the EM is constrained at high temperatures by the low-FIP \ion{Mg}{9} lines. If even higher temperature 
lines were observed, the \ion{Mg}{9} lines would add to the low-FIP diagnostic capability, but in practise their usefulness depends on how the MCMC 
solution adapts to attempt to reproduce these lines with different assumed abundances since they set the higher boundary limit of the
EM distribution. In S1 the \ion{Mg}{9} 706.02\,\AA\, line intensity is reproduced only with coronal abundances, whereas in S2 it is reproduced
with both photospheric and coronal abundances.

It is difficult with this technique to put a number on the FIP bias. We can say that the Mg enhancement factor in the adopted
elemental abundance data is a factor of 1.9, and this works. We can see in Table \ref{table2}, however, that the \ion{Mg}{8}
lines would be reproduced better if the Mg abundance were increased by a factor of 1.4.
This implies a FIP bias of 2.7; a number that
is also closer to the EIS results for the loop footpoint regions. At first sight the results for approximately the same region (EIS R4 and SPICE S2) appear
inconsistent, but are in fact in line with what we would expect based on our
understanding of the plasma compositional structure of active regions. At the temperatures of the SPICE measurements, we are detecting a strong FIP effect in the red-shifted fan loops.
At the higher temperatures accessed by EIS, we are measuring the composition of the upflows and these show a range of values. Note also that the 
measurements are made with different element pairs, and of course since the observations were taken one week apart we should also consider the evolution of plasma 
composition in decaying ARs.
Nevertheless, the combination of SPICE and EIS shows potential for allowing us to disentangle the compositional structure of active regions
at different temperatures.

An alternative is to derive the FIP bias from a minimization of the chi squared value for the \ion{Mg}{8} lines. 
This results in a similar range of values (2.6--2.7). Our analysis demonstrates that determining whether a solar
feature has a photospheric or coronal composition is possible with SPICE observations. Making a more accurate measurement
of the FIP bias is more challenging. This can also be achieved with simple techniques like the chi squared minimization we mention here, but of course results in uncertainties (40\% in our case). 
The EIS observations show that the FIP bias in the upflow regions is smaller than at the loop footpoints. This is in line with our expectations developed from studies
of the upflows/loop footpoints as potential slow wind/SEP sources. The difference for AR 12781 is around a factor of 2, which is potentially large enough to be detectable by SPICE with the methods we have employed. This is encouraging for SPICE connection studies going forward.

There is some evidence of fractionation between Si and Fe
in our EIS measurements. For the upflow regions, the scaling factor we needed to apply to the Fe-only EM to reproduce the \ion{Si}{10} 258.375\,\AA\, line
is around the 30--40\% level. This is close to the EIS calibration uncertainty for an intensity ratio. For the loop footpoints, however, the
scaling was significantly larger (a factor of 1.8--1.9). It is tempting to suggest that Si and Fe are fractionating in the closed magnetic field of
the loop footpoints, but not on the open field of the upflows. In the ponderomotive force model of the FIP effect
the Fe/Si ratio is close to 1.0 for the full range of slow-mode wave amplitudes on open magnetic field \citep[Table 4,][]{Laming2015}. In contrast, 
the Fe/Si ratio reaches 1.9 at the high end of wave amplitudes on closed field \citep[Table3,][]{Laming2015}. Of course we cannot rule out
the possibility of cross-detector calibration uncertainties. Most of the Fe lines used for the EM analysis fall on the short-wavelength (SW) detector whereas the
critical Si line falls on the long-wavelength (LW) detector. We have already noted a discrepancy for the \ion{Fe}{14} lines on the LW detector.

\section{Conclusion}

In this initial study, we combined Hinode/EIS observations of AR 12781 taken from an Earth facing view, with Solar Orbiter/SPICE 
measurements obtained one week later from a longitude of 120$^{\circ}$ West compared to Earth. The aim was to develop a plasma composition measurement technique for use with
SPICE, and benchmark it against results from a well tested DEM method routinely used to obtain the Si/S abundance ratio with EIS.

Using the EIS measurements, we characterize the general compositional structure of the AR, to ensure that it is a typical region.
As expected from previous work, the strongest FIP effect is seen in the loops in the core of the AR, especially at the footpoints. Outflows on the eastern side
also show a coronal composition, but on the western side they show a photospheric composition, which is relatively rare.

Lacking a robust abundance diagnostic ratio with a good temperature overlap between lines from high- and low-FIP elements for SPICE, 
we looked solely for consistency between observed
and DEM predicted intensities under the assumption of photospheric or coronal abundances. Measurements of the bright fan structures on the western side are
consistent with a coronal Mg/Ne composition, and again, the results suggest a stronger FIP effect at the loop footpoints. These results demonstrate
that SPICE can be used to detect coronal Mg/Ne abundances.

Based on previous work \citep{Warren2011}, the fan loops are expected to show red-shifts at the temperatures accessed by SPICE ($\sim$ 0.56\,MK),
and those structures are embedded below the blue-shifted outflows EIS observed previously at higher temperatures ($\sim$ 0.9\,MK). The fan loops have a coronal
Mg/Ne composition, consistent with earlier work, with the outflows above showing a photospheric composition. Although these observations were not
taken simultaneously, this mixed, variable composition from an AR edge that shows outflows has been identified previously as a feature that could
explain the variability in composition observed in the solar wind \citep{Brooks2020}. 
This analysis demonstrates the power of EIS and SPICE to work 
in combination to disentangle the observed emission contributions from different structures with different elemental abundances in an area of AR outflows;
a capability with potential for use in other situations, discussed in the introduction, such as the detection of the mixed composition in the core of 
ICMEs and coronal abundances 
in the surrounding plasma.

Our study has been based on establishing the general compositional structure of an AR with EIS and inferring consistency in subsequent observations
with SPICE. Moving forward we
aim to use both simultaneous measurements and complementary multi-viewpoint observations to
understand the mean magnitude and degree of variability of coronal/photospheric abundance ratios for different element pairs at diferent temperatures, so that more exact comparisons with the in situ
solar wind data can be made. The small-scale spatial distribution of abundances in active regions also needs to be established definitively, though some advances in
this direction have been made by \cite{Mihailescu2022}. Such studies
will allow both broader comparisons with composition variability in the solar wind, and direct matching between measurements with Solar Orbiter/SWA and abundance ratios in source regions identified by magnetic connectivity models.

Finally, our observations were taken during the Solar Orbiter commissioning window, when an AR fortunately passed into the FOV.
The mission is now in the nominal phase, and dedicated active region SOOPs (Solar Orbiter Observation Plans) will be made on a regular basis that
will provide a continuing understanding of the SPICE capabilities, in addition to solar wind and SEP connection science.

\begin{acknowledgments}
The work of D.H.B. and H.P.W. was funded by the NASA Hinode program. This work was supported by CNES and S.P. acknowledges the funding by CNES through the MEDOC data and operations center. D.B. is funded under STFC consolidated grant number ST/S000240/1.
Solar Orbiter is a mission of international cooperation between ESA and NASA, operated by ESA.
The development of SPICE has been funded by ESA member states and ESA. 
It was built and is operated by a multi-national consortium of research 
institutes supported by their respective funding agencies:  STFC RAL 
(UKSA, hardware lead), IAS (CNES, operations lead), GSFC (NASA), MPS 
(DLR), PMOD/WRC (Swiss Space Office), SwRI (NASA), UiO (Norwegian Space 
Agency). The EUI images are courtesy: ESA/Solar Orbiter/EUI.   
Hinode is a Japanese mission developed and launched by ISAS/JAXA, with NAOJ 
as domestic partner and NASA and STFC (UK) as international partners. 
It is operated by these agencies in co-operation with ESA and NSC (Norway). 
\end{acknowledgments}

\end{document}